\documentclass[10pt,superscriptaddress,nofootinbib,notitlepage,twocolumn,pra]{revtex4-2}
\usepackage{amsthm,amsmath,amssymb}
\usepackage{mathrsfs}
\usepackage{appendix}
\usepackage{amstext}
\usepackage{graphicx}
\usepackage{url}
\usepackage{float}
\usepackage{bm}
\usepackage{ifthen}
\usepackage[usenames,dvipsnames]{color}
\usepackage{mathrsfs}
\usepackage[colorlinks=true, citecolor=blue, urlcolor=black]{hyperref}
\usepackage{dcolumn}
\usepackage{natbib}
\usepackage{booktabs}
\usepackage{hyperref}
\usepackage{color}
\usepackage{multirow}
\usepackage{mathtools}
\usepackage{lipsum}
\usepackage[dvipsnames]{xcolor}
\usepackage{tcolorbox}
\tcbuselibrary{most}
\newcommand{\bra}[1]{\mbox{$\left \langle #1 \right|$}}
\newcommand{\ket}[1]{\mbox{$\left| #1 \right\rangle$}}

\begin{document}
	
\title{Multi-field quantum conferencing overcomes the network capacity limit}
	
\author{Yuan-Mei Xie}\thanks{These authors contributed equally.}
\author{Yu-Shuo Lu}\thanks{These authors contributed equally.}
\affiliation{National Laboratory of Solid State Microstructures and School of Physics, Collaborative Innovation Center of Advanced Microstrucstures, Nanjing University, Nanjing, China}
\affiliation{School of Physics and Beijing Key Laboratory of Opto-electronic Functional Materials and Micro-nano Devices, Key Laboratory of Quantum State Construction and Manipulation (Ministry of Education), Renmin University of China, Beijing, China}
\author{Yao Fu}
\affiliation{Beijing National Laboratory for Condensed Matter Physics and Institute of Physics, Chinese Academy of Sciences, Beijing, China}
\author{Hua-Lei Yin}\email{hlyin@ruc.edu.cn}
\affiliation{School of Physics and Beijing Key Laboratory of Opto-electronic Functional Materials and Micro-nano Devices, Key Laboratory of Quantum State Construction and Manipulation (Ministry of Education), Renmin University of China, Beijing, China}
\affiliation{National Laboratory of Solid State Microstructures and School of Physics, Collaborative Innovation Center of Advanced Microstrucstures, Nanjing University, Nanjing, China}
\affiliation{Beijing Academy of Quantum Information Sciences, Beijing, China}
\affiliation{Yunnan Key Laboratory for Quantum Information, Yunnan University, Kunming, China}
\author{Zeng-Bing Chen}\email{zbchen@nju.edu.cn}
\affiliation{National Laboratory of Solid State Microstructures and School of Physics, Collaborative Innovation Center of Advanced Microstrucstures, Nanjing University, Nanjing, China}

\begin{abstract}
Quantum conferencing enables multiple nodes within a quantum network to share a secure  conference key for private message broadcasting. The key rate, however, is limited by the repeaterless capacity to distribute multipartite entangled states across the network.  Currently, in the finite-size regime, no feasible schemes utilizing existing experimental techniques can overcome the
fundamental rate–distance limit of quantum conferencing
in quantum networks without repeaters. Here, we propose a practical, multi-field scheme that breaks this limit, involving virtually establishing Greenberger-Horne-Zeilinger states through post-measurement coincidence matching. This proposal  features a measurement-device-independent characteristic and can directly scale to support any number of users. Simulations show that the fundamental limitation on the  conference key rate can be overcome in a reasonable running time of sending $10^{14}$ pulses. We predict that it offers an efficient design for long-distance broadcast communication in future quantum networks.
\end{abstract}
	
\maketitle
\bigskip
\noindent
\textbf{\large Introduction}\\
With the rise of digital technology, remote collaboration and online meetings have become integral to modern life, enabled by the widespread use of the internet.
Recent years have seen frequent privacy breaches online, highlighting the urgent need to upgrade to the quantum internet~\cite{Stephanie2018quantum,yin2023experimental}. This form of internet offers enhanced communication, computing, and sensing capabilities  by harnessing inherently quantum properties~\cite{wang2022twin,weng2023beating,komar2014quantum,buhrman2003distributed}.
Quantum conferencing, also known as quantum conference key agreement (QCKA)~\cite{bose1998multipartie,chen2007multi,Murta2020Quantum,Cao2021coherent,cao2021high,alasdair2022continuous}, enables  multiple users to share secure  conference  keys simultaneously, facilitating  essential tasks like remote collaboration and online meetings.  From the long-term perspective of quantum networks, the development of QCKA is expected to drive innovations in experimental technologies ~\cite{gaertner2007experimental,erven2014experimental,massimiliano2021experimental,shen2023experimental,liu2023experimental,pickston2023conference,webb2024experimental}  and theoretical tools~\cite{epping2017multi,Grasselli2018finite,kaur2020fundamental,Hahn2020Anonymous,horodecki2022fundamental,li2023breaking,grasselli2022anonymous,philip2023multipartite,bao2024efficient}.

Multipartite entanglement, specifically the Greenberger-Horne-Zeilinger (GHZ) state~\cite{greenberger1989bell}, plays a key role in the security proofs of QCKA due to its characteristics as a maximally entangled state~\cite{maneva2002improved,chen2007multi}.  
Protocols based on shared multipartite entanglement have been extended to guarantee anonymity in larger networks~\cite{Hahn2020Anonymous,grasselli2022anonymous} and applied to device-independent QCKA~\cite{philip2023multipartite}. However, the
direct distribution of GHZ states over long distances poses an experimental challenge with current technology, attributed to their fragility and high susceptibility to photon losses~\cite{erven2014experimental}.  Fortunately, real entanglement resources are not essential for the practical implementation of QCKA.  A measurement-device-independent (MDI) QCKA protocol based on the post-selected GHZ state  was proposed~\cite{fu2015long}, thereby avoiding the need to prepare entanglement beforehand. Together with the decoy-state method~\cite{hwang2003decoy,lo2005decoy,wang2005decoy}, this scheme can defeat photon-number-splitting attacks~\cite{Limitations2000brassard} and be implemented over distances exceeding 100 km using conventional weak coherent pulse (WCP). Furthermore, by combining the concept of twin-field QKD~\cite{lucamarini2018overcoming} with post-selected GHZ states, phase-matching QCKA~\cite{Zhaophase2020}  was proposed to further improve the scalability of the conference  key rate. However, overcoming the fundamental limits on  conference  key rates in arbitrary quantum networks, quantified by the multipartite private state distribution capacity~\cite{pirandola2020general,Das2021Universal}, remains a formidable but important challenge without  quantum repeaters.  

Despite efforts to overcome this barrier through techniques such as spatial multiplexing with adaptive operations~\cite{li2023breaking} and single-photon interference with post-selected $W$ states~\cite{Grasselli2019conference}, advanced technical solutions remain unachievable. To tackle the issue of nonclassical sources~\cite{Grasselli2019conference}, a WCP scheme  was  proposed to post-select the $W$ state for establishing a  conference  key, where multiple laser pulses interference in a complex, customized multiport beam splitter network~\cite{Carrara2023Overcoming}. However, 
this proposal only surpasses the capacity limits in the asymptotic limit, and its security against coherent attacks in the finite-size regime remains challenging to resolve.  
We should be aware that the  effect of sending a finite number of pulses is non-negligible when implementing quantum communication protocols in practice, as monitoring an adversary's act consumes a fraction of the time and cost.
Additionally, modifications to the number of users necessitates changes to the multiport beam splitter network~\cite{Carrara2023Overcoming}, which severely restricts the flexibility to add or remove users in quantum networks.

\begin{figure}[tbp!]
	\includegraphics[width=8.5cm]{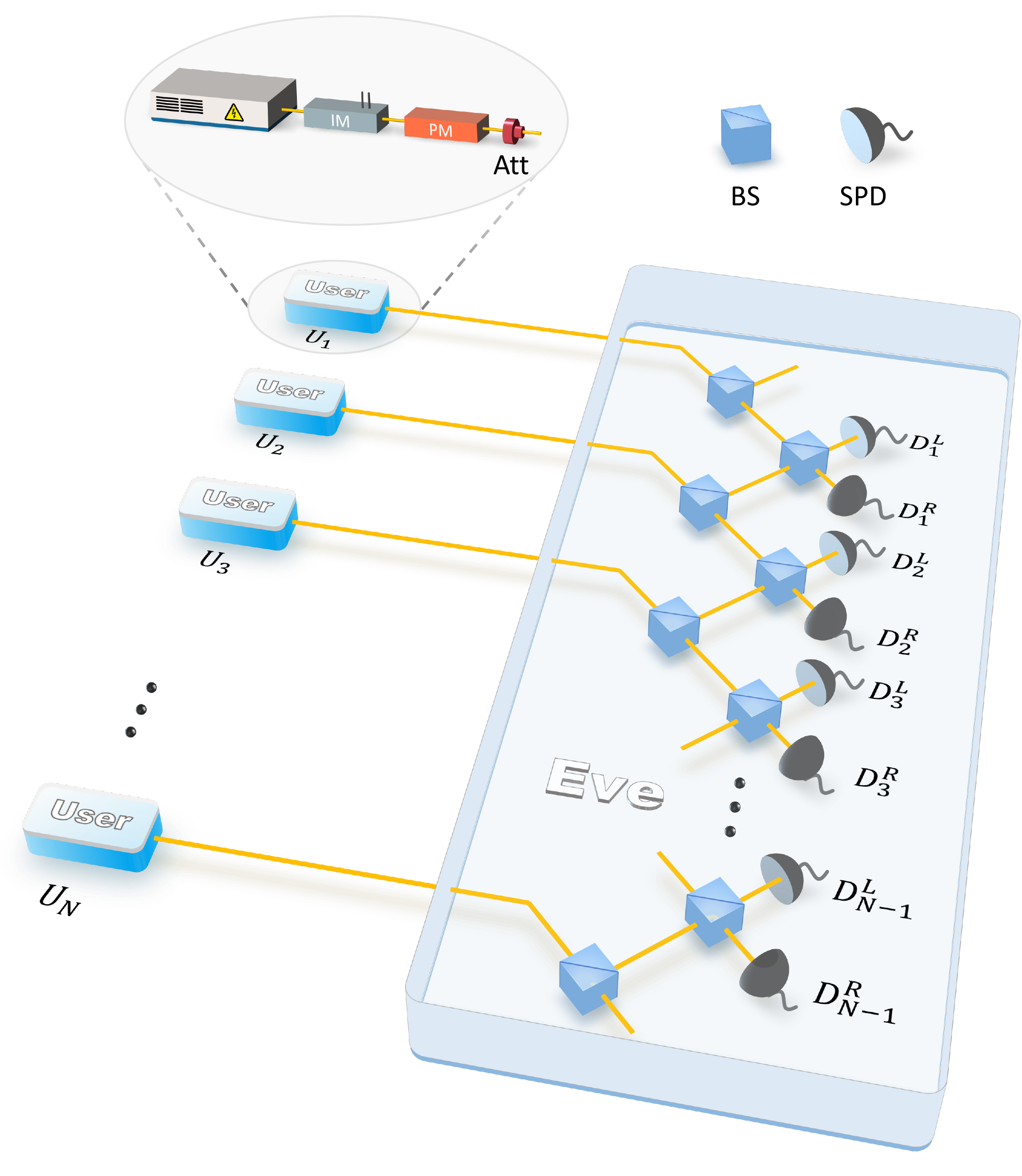}
\caption{Setup to implement MF-QCKA. 
Our setup features a network structure that can be scaled to accommodate any number of users. At each user node, a narrow-linewidth continuous-wave laser serves as the light source. Encoding is achieved through the utilization of an intensity modulator (IM) and a phase modulator (PM). These encoded light pulses are then attenuated to the single-photon level using an attenuator (Att) before being transmitted to the untrusted central measuring station, Eve, via a quantum channel. At Eve's station, incoming pulses from all users are split into two sub-pulses using beam splitters (BS). The  pulses   from one end of the BS are then subjected to interference measurements with those from the adjacent user, employing additional BS and single-photon detectors (SPDs).
  }\label{qccprotocol}
\end{figure}

 \begin{figure*}[tbp!]
\includegraphics[width=17.6cm]{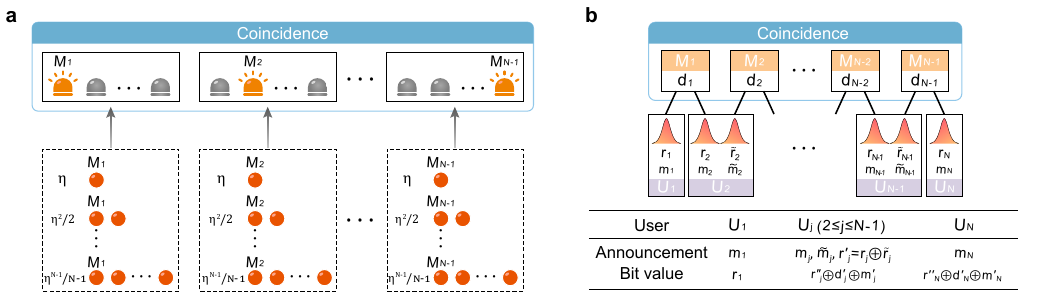}
\caption{ Detailed process  of the $N$-user MF-QCKA scheme.   \textbf{a}, Matching mechanism of  MF-QCKA. \(N-1\) successful time bins (black rectangular boxes) are matched to create a coincidence, with each bin recording a successful click at a unique measuring port among \(M_1, M_2, \ldots, M_{N-1}\). Each bin contains \(N-1\) potential click scenarios. Specifically, for a bin that records a successful click (red ball) at measuring node \(M_j\), it is possible that 1 to \(N-2\) additional measuring nodes also achieve successful clicks. The corresponding probability of selecting node \(M_j\) over others is indicated nearby. $\eta$ is the  transmittance of the quantum channel between the user and the untrusted relay.  post-measurement matching suggests that successful coincidence probability is directly proportional to the success probability of the time bin, which is of order $O(\eta)$. Consequently, the resulting key rate of our scheme scales linearly with the transmittance $\eta$.
\textbf{b}, Illustration of bit extraction within each coincidence. 
$r_j$ ($m_j$) and $\tilde{r}_j$ ($\tilde{m}_j$)  represent the random bit (phase computational results) corresponding to user \( U_j \) (\(2\leq j\leq N-1\)) in the matched time bins containing successful measuring ports \( M_{j-1}\) and \( M_j\), respectively. In each coincidence, users \(U_1\) and \(U_N\) publish their respective phase computational results   \(m_1\) and \(m_N\), while users \((U_j)_{j=2}^{N-1}\) publish their phase computational results   \(m_j\) and \(\tilde{m}_j\), along with their corresponding random bit XOR value \(r^\prime_j=r_j\oplus \tilde{r}_j\). Additionally, Eve broadcasts the corresponding detector results at the \(j\)-th measuring port as \(d_j\), where bit 0 (1) signifies a left (right) detector click. The bit value of each users is shown in the table,  where $d^\prime_j=\bigoplus_{k=1}^{j-1} d_k$, $r^{\prime\prime}_2=r_2$, $m^{\prime }_2=m_1\oplus m_2$, and for $3\le j\le N$, $r^{\prime\prime}_j=r_j\oplus_{k=2}^{j-1} r^\prime_k$ and $m^{\prime }_j=\oplus_{k=1}^{j} m_k\oplus_{k=2}^{j-1} \tilde{m}_k$.
For   derivations of the bit value, see
Methods section.		}\label{qccprotocol2}
	\end{figure*}

Here we propose a multi-field (MF) QCKA protocol based on post-selected GHZ states that can overcome the capacity limits in quantum networks in the finite-size regime. Equipped with a simple setup structure, the MF-QCKA network allows users to flexibly add or move using off-the-shelf optical devices: lasers, linear optical elements, and photodetectors.
Our protocol involves multiple fields interfering with one another at a potentially malicious central station,  where legitimate coincidences are efficiently generated through matching phase-correlated single-photon detection events using post-measurement matching method~\cite{xie2022breaking,zeng2022mode,Xie2023scalable}. 
This method ensures that the probability of a successful coincidence is proportional to the success probability of single-photon interference.
Consequently, MF-QCKA maintains the MDI characteristic while achieving a key rate that linearly depends on channel transmittance, regardless of the number of users.  Our method provides a viable structure for generating robust GHZ states even under high-loss scenarios in the network.

\bigskip
\noindent
\textbf{\large Results}\\
\textbf{MF-QCKA Protocol}\\
We introduce our MF-QCKA protocol  to distill   secure $N$-user  conference key bits using the setup shown in Fig.~\ref{qccprotocol}. Each user $(U_j)_{j=1}^N$ employs a laser, intensity  modulator, phase modulator, and attenuator to generates phase-randomized  WCPs. The signals are then sent to an untrusted relay, Eve, which comprises $N-1$ measuring ports,  denoted as $M_1, M_2, \cdots M_{N-1}$. Each measuring port is equipped with a beam splitter (BS) and two single-photon detectors.  The detectors at the $j$-th port $M_j$ are labeled as $D_j^L$ and $D_j^R$.
The incoming pulses are split into two sub-pulses using BS (labeled BS0 for differentiation) at Eve’s station. At the $j$-th measuring port, the sub-pulse from user $U_j$
(right output of a BS0) and the sub-pulse from user $U_{j+1}$ (left output of another BS0)  are subjected to  interference measurements. 
Note that to maintain the high visibility of interference, the optical phase among all the users  should be perfectly locked. 
This can be achieved either  with  passive auto-compensation techniques such as a Sagnac loop~\cite{zhong2019proof,cao2024Experimental} or  with  active global phase locking~\cite{Liu2023Experimental1000} and post-selection~\cite{ma2018phase}.

We define a click at a measuring port as successful if only one detector at the port clicks. To represent a successful click at the  measuring port $M_j$, where users $U_j$ and $U_{j+1}$ respectively send pulse intensities $k_j$ and $k_{j+1}$, we employ the notation $(k_j|k_{j+1})_j$. The variable $d_j$ is assigned the value 0 (or 1) corresponding to a click from detector $D_j^L$ (or $D_j^R$). A `successful time bin' is defined as a time bin that contains one or more successful clicks.
 The execution of the $N$-user MF-QCKA protocol involves six steps, which are summarized below. 

\textbf{Signal preparation:}
For each time bin $i=1,2,\dots, \mathcal{N}$, user  $(U_j)_{j=1}^N$  randomly generates a  weak coherent pulse $\ket{e^{i(\theta_j+r_j\pi)}\sqrt{k_j}}$, where $r_j\in\{0,1\}$ represents a random bit,   $\theta_j=2\pi \mathcal{M}_j/\mathcal{M}$ is a random phase with $\mathcal{M}_j\in\{0,1,\dots,\mathcal{M}-1\}$,   and  $k_j\in\{\mu,\mu_1,\mu_2,\dots,\mu_{N-1},0\}$ is a random intensity with probability $p_{k_j}$.  All users send their pulses to Eve via insecure quantum channels. 

 \textbf{Detection:} Eve separates every pulse from all users into two sub-pulses using BS0 for  WCP  interference measurements. 
In each successful time bin, Eve randomly selects a successful click and broadcasts the corresponding measuring port's serial number, $j$,  along with $d_j$.

\textbf{Click filtering:}   For a successful time bin, where Eve broadcasts the $j$-th measuring port, users $U_j$ and $U_{j+1}$ announce  the their preparation intensities  $k_j$, $k_{j+1}$,  phase-slice indexes   $\mathcal{M}_j,\mathcal{M}_{j+1}$.  The phase computational values $m_j$ and $m_{j+1}$ can be obtained as $m_j:=\lfloor\frac{2\mathcal{M}_j}{\mathcal{M}}\rfloor$ and $m_{j+1}:=\lfloor\frac{2\mathcal{M}_{j+1}}{\mathcal{M}}\rfloor$, respectively. 
The time bin is retained if it meets both conditions: $k_j = k_{j+1}$ and $\mathcal{M}j \equiv \mathcal{M}{j+1}\pmod{\frac{\mathcal{M}}{2}}$. Otherwise,  it is discarded. The retained successful time bins are then  organized into distinct sets $\mathcal{T}_{j}^{m}$ by calculating $m = \mathcal{M}_j \mod \frac{\mathcal{M}}{2}$.

\textbf{Coincidence matching:  } 
 For each $m=0,1,\dots,  \frac{\mathcal{M}}{2}-1$,   one time bin is randomly selected from each of the $N-1$ sets, $\mathcal{T}_{1}^{m}, \mathcal{T}_{2}^{m},\dots, \mathcal{T}_{N-1}^{m}$ to create a coincidence $[(k_1|k_1)_1,(k_2|k_2)_2,\dots,(k_{N-1}|k_{N-1})_{N-1}]$,  as schematized in Fig.~\ref{qccprotocol2}a.  The coincidence is retained if $k_1=k_2=\dots=k_{N-1}$.
  
\textbf{Sifting:}    The retained coincidences $[(k|k)_1,(k|k)_2,\dots,(k|k)_{N-1}]$ are organized  into sets $\mathcal{S}_{k}$,  and the corresponding occurrences  $s_{k}$ are  counted for  $k\in\{\mu,\mu_1,\mu_2,\dots,\mu_{N-1},0\}$.   
For coincidences in set $\mathcal{S}_{\mu}$,
 users $(U_j)_{j=2}^{N-1}$  publish their corresponding random bit XOR value $r^\prime_j=r_j\oplus \tilde{r}_j$. Here, $\tilde{r}_j$ denotes the notation corresponding to user $(U_j)_{j=2}^{N-1}$ in the click $(\mu|\mu)_j$.  These announcements  facilitates bit extraction within each coincidence (see Fig.~\ref{qccprotocol2}b).

\textbf{Parameter estimation and postprocessing:} 
 All users apply $s_{\mu}$ random bits from $\mathcal{S}_{\mu}$ to form the raw key, and utilize $s_{k}$ to estimate an upper bound  on the phase error rate, $\overline{\phi}_{\mu}^z$.
By applying error correction and privacy amplification, the key rate of $\varepsilon_{\rm{EC}}$-correct and $\varepsilon_{\rm{sec}}$-secret MF-QCKA in the finite-key regime   is given by~\cite{li2023breaking} 
	\begin{equation}\label{QCKAlength}
 \begin{aligned}
		R=&\frac{s_{\mu}}{\mathcal{N}} \left[1-H_2(\overline{\phi}_{\mu}^z)-f\times\max\limits_{j\geq 2}  H_2(E^{1,j}_{\mu})\right] \\
 & -\frac{1}{\mathcal{N}} \log_2{\frac{2(N-1)}{\varepsilon_{\rm EC}}}
  -\frac{2}{\mathcal{N}} \log_2\frac{1}{2 \varepsilon _{\rm PA}},
  \end{aligned}
	\end{equation}
where $H_2(x)=-x\log_2x-(1-x)\log_2(1-x)$ is the binary Shannon entropy function;  $f$  is the error correction efficiency and
$E_{\mu}^{1,j},j=2,3,\dots,N$, are marginal error rates, which describe the bit-flip error rates between users $U_1$ and $U_j$;  $\varepsilon _{\rm PA}$ is positive constants proportional to $\varepsilon_{\rm{sec}}$.

\begin{figure}[tbp!]
	\includegraphics[width=8.5cm]{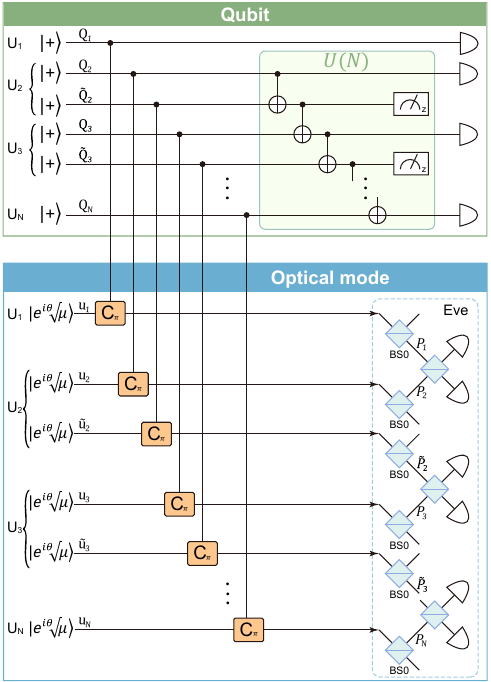}
\caption{Schematic diagram of the  virtual entanglement-based $N$-user MF-QCKA protocol.  Users create their respective entangled states by applying control phase gates $C_\pi$ to their optical modes, each in the state $\ket{e^{i\theta}\sqrt{\mu}}$, with each corresponding qubit initiated in the state $\ket{+}$.
Subsequently, the users apply operator $U(N)$, which consists of $2(N-2)$ CNOT gates,  followed by quantum measurements on the non-participatory qubits $(\tilde{Q}_j)_{j=2}^{N-1}$  in the $Z$ basis. Once the detection results of optical modes at Eve's station constitute a coincidence event, the $N$-qubit GHZ state in qubits $\{Q_1,Q_2,\cdots,Q_N\}$ can be obtained. }\label{qcka_proof}
\end{figure}

In the MF-QCKA protocol, the success of a time bin depends on the occurrence of successful single-photon interference from at least one measuring port on the relay, with a probability of $O(\eta)$. Here, $\eta$ represents the transmittance of the quantum channels linking each user to Eve (assuming symmetric channels).  The post-matching process makes $N-1$ successful time bins form a coincidence, which merely adds a factor of $1/(N-1)$ to the $O(\eta)$ scaling for the asymptotic key rate. In contrast, the phase-matching QCKA protocol~\cite{Zhaophase2020} employs a nearly identical setup for postselecting GHZ states and scales with $O(\eta^{N-1})$. This scaling is due to its predetermined matches, which are independent of single-photon interference success.  Since the $N$-user MF-QCKA protocol uses $2(N-1)$ detectors, the signal-to-noise ratio  is reduced by a factor of $2(N-1)$, caused by the dark count rate $p_d$  of each individual detector. 

\bigskip
\noindent
\textbf{Security analysis}\\
We introduce a virtual entanglement-based MF-QKD protocol, as illustrated in Fig.~\ref{qcka_proof}, which  elucidates the core ideas that lead to the MF-QCKA protocol. In the virtual protocol, each user prepares an entangled state between their virtual qubit and a WCP,  instead of directly preparing a WCP in each time bin.
Users pre-match the $t_1$-th, $t_2$-th, $\cdots$, $t_{N-1}$-th time bins. Without loss of generality, we assume uniform phases among all users (\(\theta\)) and uniform intensity across all users (\(\mu\)). We use $Q_1$~$(Q_N)$ and $u_1$~$(u_N)$ to  represent the qubit and optical mode  of user $U_1$  in the $t_{1}$-th ($t_N$-th) time bin, respectively.
For $2\le j\le N-1$,  $Q_j$ and $\tilde{Q}_j$ respectively represent the qubits of user $U_j$  in the \(t_{j-1}\)-th and \(t_j\)-th time bins,  while  $u_j$ and $\tilde{u}_j$ respectively represent the optical modes of user $U_j$  in the \(t_{j-1}\)-th and \(t_j\)-th time bins.
To prepare entangled states, users first generate composite states, each consisting of a WCP and an qubit initialized in $\ket{+}$. Next, entanglement is generated by applying control gates $C_\pi=\ket{0}\bra{0} U_0+\ket{1}\bra{1} U_\pi$ to their qubits and optical modes, where $U_{0(\pi)}$  imparts a $0(\pi)$ phase shift to the optical mode. Following these operations, the joint quantum state of the system is given by
 \begin{equation}
\begin{aligned}
&\frac{1}{2^{N-1}}  \bigotimes_{j=1}^N\left(\sum_{r_j\in\{0,1\}}\ket{r_j}_{Q_j}\ket{e^{i[\theta+(m_j+r_j)\pi]}\sqrt{\mu}}_{u_j}\right)\\
&\bigotimes_{j=2}^{N-1}\left(\sum_{\tilde{r}_j\in\{0,1\}}\ket{\tilde{r}_j}_{\tilde{Q}_j}\ket{e^{i[\theta+(\tilde{m}_j+\tilde{r}_j\pi)]}\sqrt{\mu}}_{\tilde{u}_j}\right).
 \end{aligned}
 \end{equation}
where  $m_j,\tilde{m_j}$ are random bit values indicating $0/\pi$-phase.

According to the sequence in time bins, users keep  their qubits in quantum memory and send optical modes to Eve via insecure quantum channels. Subsequently, the users apply the operator $U(N)$, which corresponds to  a series of CNOT gates,  followed by quantum measurements on non-participatory qubits $(\tilde{Q}_j)_{j=2}^{N-1}$  in the $Z$ basis.
After the optical modes passes through BS0 and the qubits undergoes  the $U(N)$ operation, the joint quantum state evolves to (this deduction process is detailed in the Methods section)
\begin{equation}
\begin{aligned}
& \frac{1}{2^{(2N-3)/2}}\times\sum_{i_2,\tilde{i}_2,\cdots,i_{N}\in\{0,1\}}\\
&\left\{\frac{1}{\sqrt{2}}\left[\ket{0i_2l_3\cdots l_N}+\ket{1\overline{i}_2\overline{l}_3\cdots\overline{l}_N}\right]\sqrt{p_{\rm even}}\ket{\Phi_{\rm even}}\right.\\
&\left.+\frac{1}{\sqrt{2}}\left[\ket{0i_2l_3\cdots l_N}-\ket{1\overline{i}_2\overline{l}_3\cdots\overline{l}_N}\right]\sqrt{p_{\rm odd}}\ket{\Phi_{\rm  odd}}\right\},
 \end{aligned}
 \end{equation}
 where  $l_j=i_j\oplus_{k=2}^{j-1}(i_k\oplus \tilde{i}_k)$, and $\ket{\Phi_{\rm even(odd)}}$ is defined in Eq.~\eqref{eq9}.  
 Once these $N-1$ time bins form a successful coincidence, the remaining qubits  $\{Q_1,Q_2,\cdots,Q_N\}$  will establish entanglement:
\begin{equation}
\begin{aligned}
\ket{\Phi_{\rm GHZ}}=\frac{1}{\sqrt{2}}\left[\ket{0i_2l_3\cdots l_N}+(-1)^{n}\ket{1\overline{i}_2\overline{l}_3\cdots\overline{l}_N}\right],
 \end{aligned}\label{eq4}
 \end{equation}
where $n=n_1+n_2+\cdots+n_{N-1}$ denotes the scenario where the joint state emits $n$ photons.
With use of the entanglement-distillation protocol~\cite{maneva2002improved}, one can distill the perfect $N$-qubit GHZ state, $\ket{\Phi^+}=\frac{1}{\sqrt{2}}\left(\ket{0 }^{\otimes N}+\ket{1}^{\otimes N}\right)$, to generate secret-key bits~\cite{chen2007multi,fu2015long}. We remark that the matching time bins are predetermined in the above virtual protocol. However, to address the large transmission loss of optical modes, users can perform post-matching to increase the number of valid coincidences. The security does not change since we only consider the successful matching results, and other cases are ruled out. 
 
For each of these $N-1$ time bins, the probability of clicks responding from the vacuum and multi-photon components is much smaller than the probability of clicks from the single-photon component. So we define  the phase error correlation:
\begin{equation}
E_x=\begin{cases}
0, &\mbox{if } n \equiv  N-1\pmod{2},\\
1, &\mbox{if } n \equiv  N\pmod{2}.
\end{cases}
 \end{equation}
Based on the  above definition of error under $X$ basis, we can express the phase error rate as
\begin{equation}
\phi_\mu^z=\begin{cases}
\frac{\sum_{n=0}^\infty s_{\mu}^{2n}}{s_{\mu}}, &\mbox{if } N\in \rm even,\\
\frac{\sum_{n=0}^\infty s_{\mu}^{2n+1}}{s_{\mu}}, &\mbox{if } N\in \rm odd.
\end{cases}
 \end{equation}
where $s_{\mu}^n$ represents  the number of contributions to coincidences in set $\mathcal{S}_\mu$ when the total number of photons emitted is $n$. Methods for deriving analytical upper bounds on the phase error rate using the multiparty decoy-state method are detailed in Supplementary Note 2. Additionally, an explicit simulation formula for $s_{\mu}^n$ in lossy channels is provided in Supplementary Note 3. 

\begin{figure}[t!]
		\includegraphics[width=8.5cm]{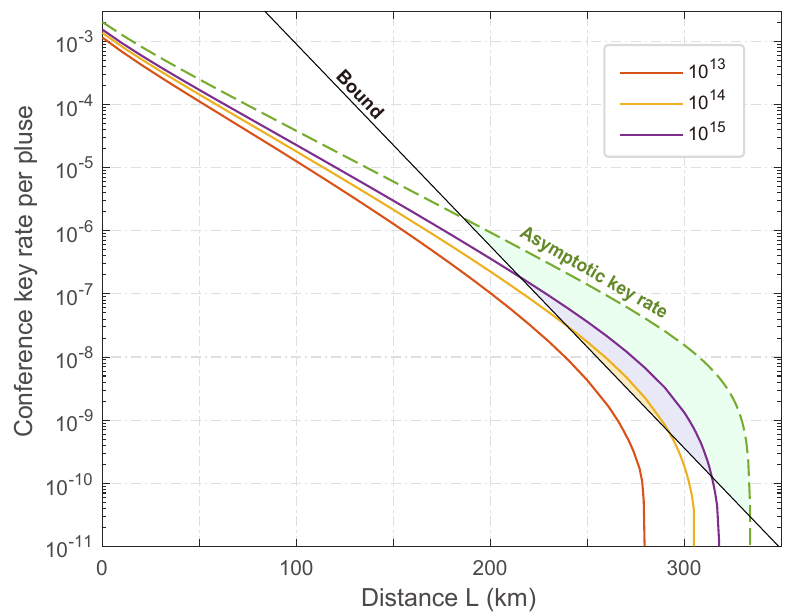}
		\caption{ Conference  key rate of 3-user MF-QCKA with three decoy states versus  distance $L$.  Here,  $L$ denotes the distance between each user and the measuring station.  The parameters used in the simulation is specified in the Methods section. Three solid colored lines correspond to different values for the total number of signals sent by users: $10^{13}$, $10^{14}$, and $10^{15}$.  Additionally, the green dotted line represents asymptotic rates with three decoy states, while the black solid line indicates the single-message multicast bound for the star network, $-\log_2(1-\eta^2)$~\cite{pirandola2020general}.  The shaded regions highlight the scenarios where increasing the number of total signals leads to improved performance. 
  Our results showcase the protocol's ability to overcome the multicast bound, even with a realistic finite size of data.
		}\label{qcka_dis_finite}
	\end{figure}

 \bigskip
\noindent
\textbf{Numerical simulation}\\
In the following, we analyze the  the performance of our protocol. In our simulation,  we consider a symmetric structure where each user is equidistant from the measuring station, with the distance denoted as $L$, and the optical signal intensities are independent of the users. The experimental parameters used in the numerical simulation is presented in Methods.  Detailed calculations for the number simulations are provided in Supplementary Note 1.
We optimize the group of parameters \((\mu, \mu_1, \cdots, \mu_{N-1}, p_\mu, p_{\mu_1}, \cdots, p_{\mu_{N-1}})\) to maximize the  conference key rate for  \(N\)-user MF-QCKA at each distance.

Figure \ref{qcka_dis_finite} illustrates the final key rates of 3-user MF-QCKA with three decoy states when the total number of signals $\mathcal{N}=10^{13},~10^{14},~10^{15}$ and in the case of infinite size.  
 To evaluate the performance of our protocol in a multipartite context, we compare its conference key rate to the maximum achievable conference key rate  in a quantum network without a relay.
This comparison is based on the single-message multicast bound of the quantum network~\cite{pirandola2020general},  which depends on the network architecture. In our scenario—a star network configuration—there is a pure-loss bosonic channel with transmittance $\eta^2$ between user $U_1$  and each other user $(U_j)_{j=2}^N$, where $\eta =10^{-\alpha L/10}$ and $\alpha$ is the attenuation coefficient of the fiber.  In this case, the single-message multicast bound is independent of the number of  users $N$ and is given by~\cite{pirandola2020general}:  $-\log_2(1- \eta^2)$.  We observe that our protocol can reach distances of over 270 km, 300 km, and 310 km when 
$\mathcal{N}=10^{13},10^{14},10^{15}$, respectively.  
The most striking feature of Fig.~\ref{qcka_dis_finite} is that the 3-user MF-QCKA scheme can exceed the bound at large distances, even in the finite-key regime, which stems from the post-measurement matching approach. At $\mathcal{N}=10^{14}$, it barely surpasses the
bound, and at $\mathcal{N}=10^{15}$, it clearly beats the bound at approximately 250 km. At a fiber distance of 50 km and with a data size of $10^{14}$, the  conference  key rate reaches $1.44 \times 10^{-4}$ bits per pulse. Assuming a 4-GHz clock rate and a 40\% duty ratio~\cite{wang2022twin}, the key rate can achieve 230.4 kbit per second. This rate is sufficient to support real-time encryption for tasks such as audio conference calls using standard symmetric encryption techniques.

\begin{figure}[t!]
\includegraphics[width=8.5cm]{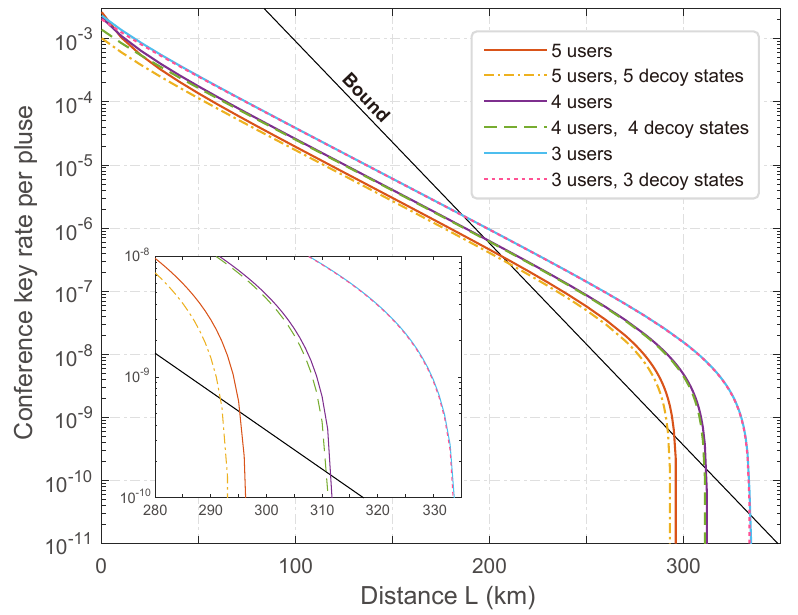}
\caption{ Asymptotic conference key rates of MF-QCKA per pulse versus distance $L$ for different numbers of users.  The experimental parameters are the same as those in Fig.~\ref{qcka_dis_finite}. Solid (dashed) lines denote a lower bound on the secret key rate with users employing an infinite (finite) number of decoy settings.  The black solid line represents the single-message multicast bound, $-\log_2(1-\eta^2)$. The inset focuses on the key rate behavior in the range 280–335 km, showing how the finite and infinite decoy state curves align closely. This alignment suggests that the finite number of decoy states effectively bounds the key rate.} \label{figqcc}
\end{figure}

To assess the potential of our protocol, we analyze its performance across varying numbers of users in the asymptotic regime. Figure~\ref{figqcc} displays the asymptotic key rates of MF-QCKA for $N=3$, $N=4$, and $N=5$ users, comparing results obtained with infinite decoy states (solid lines) and finite decoy states (dashed lines).  We remark that  the key rates with infinite decoy states for 5 users within 10 km are slightly higher than those for 4 users due to inaccuracies in the analytical estimation procedure for obtaining $s_\mu^n$ in extremely low-loss regimes (see Supplementary Note 3 for more details). We can see that our protocol can surpass the  bound for  3, 4 and 5 users in the asymptotic scenario. 
Additionally, the subtle  enhancement in key rates under high-loss conditions with infinite decoy states suggests that  a finite number of decoy states can effectively estimate the yield and phase error rate. Our results  demonstrate the achievement of a transmission distance exceeding 300 km for both 3 and 4 user, underscoring the protocol's capability for metropolitan-scale deployment.

\begin{figure}[tbp!]
\includegraphics[width=8.5cm]{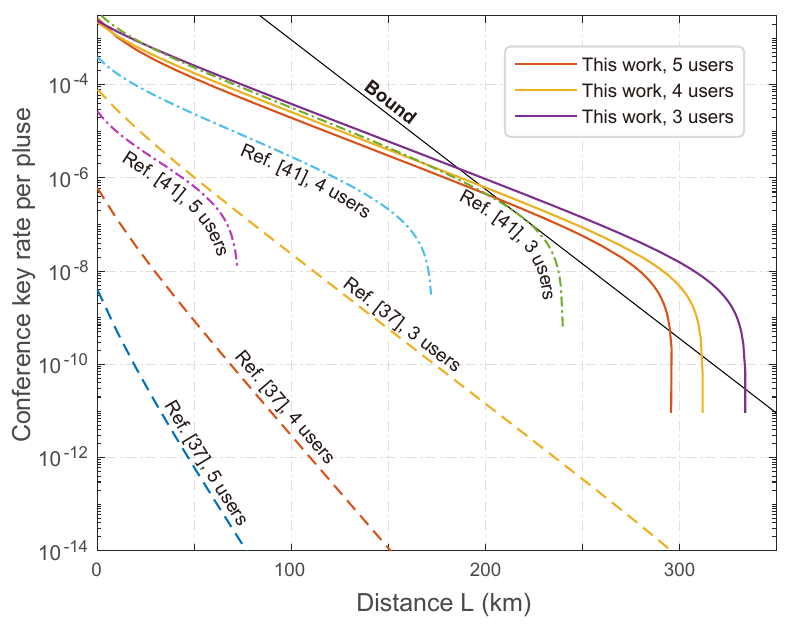}
\caption{Comparison of asymptotic key rates between MF-QCKA and other QCKA protocols with infinite decoy states versus distance $L$. The parameters used in the simulation is specified in the Methods section.  Dash-dotted lines indicate key rates of QCKA based on post-selected $W$ states~\cite{Carrara2023Overcoming}, while dashed lines represent key rates of phase-matching QCKA~\cite{Zhaophase2020}.  The black solid line denotes the single-message multicast bound, $-\log_2(1-\eta^2)$. The comparison highlights significant limitations in key-generation rate and communication distance for the latter two protocols as the number of users increases.
		}\label{figqccall}
	\end{figure}

In Fig.~\ref{figqccall}, we plot the asymptotic key rate of our  protocol alongside phase-matching QCKA~\cite{Zhaophase2020}, and QCKA based on post-selected $W$ states ($W$-state QCKA)~\cite{Carrara2023Overcoming} under $N=3,~4,~5$, using an infinite number of decoy states. We optimize the key rate of these protocols at each distance. For distances under 20 km and with 3 users, our protocol’s key rate is slightly lower than that of the $W$-state QCKA protocol. However, at greater distances and with more users, our protocol surpasses the other two in performance. The performance of  our protocol exceeds that of the phase-matching QCKA protocol by three and seven orders of magnitude at distances of 100 km for $N=3$ and $N=4$, respectively. An important property of our scheme is its robustness to an increasing number of users. Our protocol utilizes GHZ state correlations formed via post-measurement coincidence matching,  enabling key rate scaling independent of the number of users while 
maintaining a low bit error rate. In contrast, the key rate of phase-matching QCKA scales with $\eta^{N-1}$, and $W$-state QCKA suffer from intrinsic bit errors  inherited from $W$-state correlations. Consequently, the key rates of both of these protocols drop rapidly as the number of users increases.

\bigskip
\noindent
\textbf{\large Discussion}\\
In this work, we proposed a scalable multi-field QCKA protocol for distributing long-distance multi-user  conference  keys and present a comprehensive performance analysis. The simulated key rate of  3-user MF-QCKA shows that it surpasses the fundamental limitation on  conference key rates when each user sends a total of $10^{14}$ pulses, corresponding to several hours in a system with a 4 GHz pulse repetition rate. Given that  our results apply in the finite-key regime, we believe our scheme has broad practical applications, including the efficient achievement of quantum-secure conference calls~\cite{massimiliano2021experimental}. 
In the protocol, we analyzed the scenario of symmetric channel losses. Further work is  necessary to develop a more general decoy-state analysis with independent decoy intensities for each user, addressing real-life scenarios of asymmetric channel losses. 

A key element of our design is the asynchronous matching method, which forms legitimate coincidences from phase-correlated clicks. 
Owing to this method, the 4-user  conference  key rate can improve by seven orders of magnitude with essentially the same setup and technology at a distance of 100 km between each user and the measuring station.
By integrating the post-measurement matching approach with quantum memories, one can design a robust quantum repeater to generate GHZ states across the network.

 We emphasize that both laser-based and entanglement-based protocols are essential for the development of future quantum networks, each contributing unique advantages to different aspects of network architecture. Multipartite entanglement resources  offer  advantages in established quantum networks, facilitating tasks such as connecting remote quantum computers through quantum correlations. At the same time, the flexibility and accessibility of laser-based methods ensure their continued value, particularly in scenarios where entanglement resources may not yet be readily available. As quantum network architectures evolve, hybrid systems that integrate both laser-based and entanglement-based approaches are poised to play a pivotal role in enabling flexible and efficient multi-user quantum networks. 
 Building on the strengths of multipartite entanglement, 
we  anticipate near-future applications such as networks of sensors and clocks, distributed quantum computation, and genuine multipartite quantum communication, potentially leading to the development of the quantum internet.

\bigskip
\noindent
\textbf{\large Methods}\\
\textbf{Parameters of the simulation}\\	
The results of the simulations presented in Figs.~\ref{qcka_dis_finite},  \ref{figqcc} and \ref{figqccall} are obtained using the following parameters, which reflect state-of-the-art technologies~\cite{zhou2023experimental}: detector efficiency (\(\eta_{d}\)) of 77\%, a dark count rate (\(p_{d}\)) of $3.03\times10^{-9}$, error correction efficiency (\(f\)) of 1.1, an attenuation coefficient of the fiber (\(\alpha\)) of 0.16 dB/km, and the number of phase slices (\(\mathcal{M}\)) set to 16. Additionally, for Fig.~\ref{qcka_dis_finite}, we fixed the error correction failure probability (\(\varepsilon_{\rm EC}\)) at \(10^{-15}\) and the privacy amplification failure probability (\(\varepsilon_{\rm PA}\)) at \(10^{-10}\). The failure probability considered in the finite data analysis processes is \(10^{-10}\).

\bigskip
\noindent
\textbf{Quantum evaluation of the joint state}\\
After the users perform the $U(N)$ operation, the joint quantum state of the system evolves as
 \begin{equation}
\begin{aligned}
&\frac{1}{2^{N-1}}  \bigotimes_{j=1}^N\left(\sum_{r_j\in\{0,1\}}\ket{r_j}_{Q_j}\ket{e^{i[\theta+(m_j+r_j)\pi]}\sqrt{\mu}}_{u_j}\right)\\
&\bigotimes_{j=2}^{N-1}\left(\sum_{\tilde{r}_j\in\{0,1\}}\ket{\tilde{r}_j}_{\tilde{Q}_j}\ket{e^{i[\theta+(\tilde{m}_j+\tilde{r}_j\pi)]}\sqrt{\mu}}_{\tilde{u}_j}\right)\\
 =&\frac{1}{2^{N-1}} \sum_{r_1,i_2,\tilde{i}_2,\cdots,i_{N}\in\{0,1\}}
\left\{\ket{r_1}\bigotimes_{j=2}^N\ket{i_j\oplus r_1}\bigotimes_{j=2}^{N-1}\ket{\tilde{i}_j\oplus r_1}\right.\\
& \left.\bigotimes_{j=1}^N\ket{e^{i[\theta+(m_j+i_j+r_1)\pi]}\sqrt{\mu}}\bigotimes_{j=2}^{N-1}\ket{e^{i[\theta+(\tilde{m}_j+\tilde{i}_j+r_1)\pi]}\sqrt{\mu}} \right\} \\
&\xrightarrow{U(N)}\! \frac{1}{2^{N-1}}\!\sum_{r_1,i_2,\tilde{i}_2,\cdots,i_{N}\in\{0,1\}}\!
\left\{\!\ket{r_1}\ket{i_2\oplus r_1}\bigotimes_{j=3}^N\ket{l_j\oplus r_1}\right.\\
& \left.\bigotimes_{j=1}^N\ket{e^{i[\theta+(m_j+i_j+r_1)\pi]}\sqrt{\mu}}\bigotimes_{j=2}^{N-1}\ket{e^{i[\theta+(\tilde{m}_j+\tilde{i}_j+r_1)\pi]}\sqrt{\mu}} \right\} 
 \end{aligned}
 \end{equation}
 where
 \begin{equation}
\begin{aligned}
&l_j=i_j\oplus_{k=2}^{j-1}(i_k\oplus \tilde{i}_k).
 \end{aligned}\label{eq8} 
 \end{equation} 
Consider a BS characterized by the creation operators for its two output modes, denoted by 
$C_1^{\dagger}$ and $C_2^{\dagger}$. For input states $\ket{e^{i(\theta+A\pi)}\sqrt{\mu}}$ and $\ket{e^{i(\theta+B\pi)}\sqrt{\mu}}$, where $A$ and $B$ are integers, the transformation of the output state is given by:
  \begin{equation}
\begin{aligned}
&\ket{e^{i(\theta+A\pi)}\sqrt{\mu}} \ket{e^{i(\theta+B\pi)}\sqrt{\mu}}\\
\xrightarrow{\rm BS} &\frac{e^{-\mu}(\sqrt{2\mu})^n}{n!}\sum_{n=0}^\infty\left[\frac{(-1)^{A}e^{i\theta}C_1^{\dagger}+(-1)^{B}e^{i\theta}C_2^{\dagger}}{\sqrt{2}}\right]^{n}\ket{\rm vac}.
\end{aligned}\label{bs_evolve}
 \end{equation} 
Let $P^{\dagger}_1,P^{\dagger}_2,\tilde{P}^\dagger_2,\cdots, P^\dagger_N$ denote the creation operations for the corresponding path modes after BS0. By employing Eq.~\eqref{bs_evolve} in Eq.~\eqref{eq8}, we derive the output state after BS0 as follows:
  \begin{equation}
\begin{aligned}
&\frac{1}{2^{(2N-3)/2}}\times\sum_{i_2,\tilde{i}_2,\cdots,i_{N}\in\{0,1\}}\\
&\left\{\frac{1}{\sqrt{2}}\left[\ket{0i_2l_3\cdots l_N}+\ket{1\overline{i}_2\overline{l}_3\cdots\overline{l}_N}\right]\sqrt{p_{\rm even}}\ket{\Phi_{\rm even}}\right.\\
&\left.+\frac{1}{\sqrt{2}}\left[\ket{0i_2l_3\cdots l_N}-\ket{1\overline{i}_2\overline{l}_3\cdots\overline{l}_N}\right]\sqrt{p_{\rm odd}}\ket{\Phi_{\rm odd}}\right\}\\
 \end{aligned}
 \end{equation}
where 
  \begin{equation}
\begin{aligned}
&\quad \ket{\Phi_{\rm even(odd)}} =\frac{e^{-(N-1)\mu}}{\sqrt{p_{\rm even(odd)}}}\sum_{n_1+n_2+\cdots+n_{N-1}\in \rm  even(odd)}\\
&\left\{\frac{ (\sqrt{2\mu})^{ n_1+n_2+\cdots+n_{N-1}}}{   {n_1!n_2!\cdots n_{N-1}!}}\left[\frac{(-1)^{m_1}P^{\dagger}_1+(-1)^{m_2+i_2}P^{\dagger}_{2}}{\sqrt{2}}\right]^{n_1}\right.\\
&\left.\bigotimes_{j=2}^{N-1}\left[\frac{(-1)^{\tilde{m}_j+\tilde{i}_j}\tilde{P}^{\dagger}_{j}+(-1)^{m_{j+1}+ i_{j+1}}P^{\dagger}_{j+1}}{\sqrt{2}}\right]^{n_j}\right\}\ket{\rm vac},
 \end{aligned}\label{eq9}
 \end{equation} 
 and
\begin{equation}
\begin{aligned}
p_{\rm k}&=\sum_{n_1+n_2+\cdots+n_{N-1}\in \rm k}\frac{e^{-2(N-1)\mu}(2\mu)^{n_1+n_2+\cdots+n_{N-1}}}{n_1!n_2!\cdots n_{N-1}!}\\
&=\begin{cases}
e^{-2(N-1)\mu}\cosh{\left[2(N-1)\mu\right]}, & \mbox{k = even},\\
e^{-2(N-1)\mu}\sinh{\left[2(N-1)\mu\right]}, & \mbox{k = odd}.
\end{cases}
 \end{aligned}
 \end{equation} 
 Here, to simplify the notation we omit the labels $i_2,\tilde{i}_2,\cdots,i_N$ on $\ket{\Phi_{\rm even}}$ and $\ket{\Phi_{\rm odd}}$. Additionally, we exclude the phase variable $\theta$ from consideration, as Eve's strategy for announcements cannot be dependent on $\theta$. 

\bigskip
\noindent
\textbf{Derivations of the bit value}\\
 In the virtual entanglement-based $N$-user MF-QCKA protocol, shown in Fig.~\ref{qcka_proof}, $Z$-basis measurements commute with all CNOT gates. 
This allows users to equivalently advance the $Z$-basis measurement of bits
 $(Q_j)_{j=1}^N$ and $(\tilde{Q}_j)_{j=2}^{N-1}$ to the initial stage of the protocol. The measurement outcomes on the quantum bits  following the operation $U(N)$  can be equivalently obtained via classical post-processing. In this way, we reduce the entanglement-based scheme to a prepare-and-measure version. Specifically,  users $U_j$ and $U_{j+1}$ prepare weak coherent pulses $\ket{e^{i[\theta+(\tilde{m}_j+\tilde{r}_j)\pi]}\sqrt{\mu}}$ ($\ket{e^{i[\theta+(m_1+r_1)\pi]}\sqrt{\mu}}$ for $U_1$) and $\ket{e^{i[\theta+(m_{j+1}+r_{j+1})\pi]}\sqrt{\mu}}$ in the $t_j$-th time bin, respectively, Here $r_j,\tilde{r}_j$ are random bits  from the set \{0, 1\}. After Eve's measurements and announcements, 
 users \(U_1\) and \(U_N\) publish  their respective values, \(m_1\) and \(m_N\). Additionally,  users \((U_j)_{j=2}^{N-1}\) publish  \(m_j\) and \(\tilde{m}_j\), along with their corresponding random bit XOR value \(r^\prime_j=r_j\oplus \tilde{r}_j\).  By performing the corresponding series of XOR operations on the bits, the bit value for user $(U_j)_{j=2}^N$ becomes
\begin{equation}
\mathcal{R}_j=\begin{cases}
r_j, & j=2,\\
r_j\oplus_{k=2}^{j-1}(r_k\oplus\tilde{r}_k), &3\le j\le N.
\end{cases}\label{eq10}
 \end{equation} 
From the perspective of Eq.~\eqref{eq4}, once these $N-1$ time bins form a successful coincidence, the bit value of the users and user $U_1$ will satisfy the following relationship:
  \begin{equation}
\begin{aligned}
\mathcal{R}_2\oplus i_2=r_1,\qquad
\mathcal{R}_j\oplus l_j=r_1.
 \end{aligned}\label{eq11}
 \end{equation} 
From Eq.~\eqref{eq9},  we can see that  for the first measuring port,  when $m_1 = m_2 \oplus i_2$, only detector $D_1^L$ clicks. Conversely, when $m_1 \oplus 1 = m_2 \oplus i_2$, only detector $D_1^R$ clicks. For the $j$-th ($2\le j\le N$) measuring port, when $\tilde{m}_j\oplus \tilde{i}_j= m_{j+1}\oplus i_{j+1}$, only detector $D_j^L$ clicks, while when $\tilde{m}_j\oplus \tilde{i}_j\oplus 1=m_{j+1}\oplus i_{j+1}$, only detector $D_j^R$ clicks. Thus, the relationship between the measurement result published by Eve, $d_j$, the values $i_j$ and $\tilde{i}_j$, and the users' values $m_j$ and $\tilde{m}_j$, is
\begin{equation}
d_j=\begin{cases}
m_1\oplus m_2\oplus i_2, & j=1,\\
\tilde{m}_j\oplus \tilde{i}_j\oplus m_{j+1}\oplus i_{j+1}, &2\le j\le N.
\end{cases}\label{eq12}
 \end{equation} 
By combining Eq.~\eqref{eq8} and Eqs.~\eqref{eq10}-\eqref{eq12}, we establish the correct correlation of the bit values between users $(U_j)_{j=2}^N$ and $U_1$.  Specifically, the bit value should be calculated as 
  \begin{equation}
\begin{aligned}
\mathcal{R}_2\oplus i_2=&r_2\oplus d_1\oplus m_1\oplus m_2\\
\mathcal{R}_j\oplus l_j=&r_j\oplus_{k=2}^{j-1}(r_k\oplus\tilde{r}_k) \oplus i_j\oplus_{k=2}^{j-1}(i_k\oplus \tilde{i}_k)\\
=&r_j\oplus_{k=2}^{j-1}(r_k\oplus\tilde{r}_k)\oplus i_2\oplus_{k=2}^{j-1}( \tilde{i}_k\oplus i_{k+1})\\
=&r_j\oplus_{k=2}^{j-1}(r_k\oplus\tilde{r}_k)\oplus_{k=1}^{j-1}d_k \oplus_{k=1}^{j} m_k\oplus_{k=2}^{j-1} \tilde{m}_k.
 \end{aligned} \label{bit_value}
 \end{equation} 
Practically, users in the MF-QCKA scheme, described in the subsection ``MF-QCKA Protocol'' in the Results, process successful time bins after matching instead of using a predetermined match. Nevertheless, the bit extraction process remains unchanged, as it considers only the successful matching results and rules out all other cases. In this way, we conclude the proof of the key  mapping as illustrated in Fig.~\ref{qccprotocol2}b.

 \bigskip
\noindent
\textbf{\large Data availability}\\
 The data that support the findings of this study are available from the corresponding author upon reasonable request.
 
 \bigskip
\noindent
\textbf{\large Code availability}\\
The code generated for the current study are available from the corresponding author on reasonable request.

\bigskip
\noindent
\textbf{\large References}

%

\bigskip
\noindent
\textbf{\large Acknowledgments}\\
We gratefully acknowledge the supports from the National Natural Science Foundation of China (No. 12274223), the Program for Innovative Talents and Entrepreneurs in Jiangsu (No. JSSCRC2021484), the Fundamental Research Funds for the Central Universities and the Research Funds of Renmin University of China (No. 24XNKJ14). 

\bigskip
\noindent
\textbf{\large Author contributions}\\
All authors contributed to the scientific discussions and refinement of the manuscript. Z.-B.C. guided the work. H.-L.Y. conceived and supervised the research. 
 Y.-M.X., Y.-S.L., and
H.-L.Y. finished all theoretical simulation and manuscript preparation with the help of Y.F.

  \bigskip
\noindent
\textbf{\large Competing interests}\\
The authors declare no competing interests.

\end{document}


\title{Supplementary information:\\ Multi-field quantum conferencing overcomes the network capacity limit}

\author{Yuan-Mei Xie}\thanks{These authors contributed equally.}
\author{Yu-Shuo Lu}\thanks{These authors contributed equally.}
\affiliation{National Laboratory of Solid State Microstructures and School of Physics, Collaborative Innovation Center of Advanced Microstrucstures, Nanjing University, Nanjing, China}
\affiliation{School of Physics and Beijing Key Laboratory of Opto-electronic Functional Materials and Micro-nano Devices, Key Laboratory of Quantum State Construction and Manipulation (Ministry of Education), Renmin University of China, Beijing, China}
\author{Yao Fu}
\affiliation{Beijing National Laboratory for Condensed Matter Physics and Institute of Physics, Chinese Academy of Sciences, Beijing, China}
\author{Hua-Lei Yin}\email{hlyin@ruc.edu.cn}
\affiliation{School of Physics and Beijing Key Laboratory of Opto-electronic Functional Materials and Micro-nano Devices, Key Laboratory of Quantum State Construction and Manipulation (Ministry of Education), Renmin University of China, Beijing, China}
\affiliation{National Laboratory of Solid State Microstructures and School of Physics, Collaborative Innovation Center of Advanced Microstrucstures, Nanjing University, Nanjing, China}
\affiliation{Beijing Academy of Quantum Information Sciences, Beijing, China}
\affiliation{Yunnan Key Laboratory for Quantum Information, Yunnan University, Kunming, China}
\author{Zeng-Bing Chen}\email{zbchen@nju.edu.cn}
\affiliation{National Laboratory of Solid State Microstructures and School of Physics, Collaborative Innovation Center of Advanced Microstrucstures, Nanjing University, Nanjing, China}

\maketitle

\tableofcontents

\clearpage

 \section{Conference key rate calculation}
\subsection{Key rate formula} 
We denote $\underline{x}$ and $\overline{x}$ as the lower and upper bounds of the observed value $x$, respectively. The  conference key rate $R$ of the $N$-user MF-QCKA protocol  in the finite-size regime can be written as
\begin{align}
R=&\frac{s_{\mu}}{\mathcal{N}} \left[1-H_2(\overline{\phi}_{\mu}^z)-f\times\max\limits_{j\geq 2}  H_2(E^{1,j}_{\mu})\right]   -\frac{1}{\mathcal{N}} \log_2{\frac{2(N-1)}{\varepsilon_{\rm EC}}}
  -\frac{2}{\mathcal{N}} \log_2\frac{1}{2 \varepsilon _{\rm PA}},\label{eq_keyrate}
\end{align}
where $\mathcal{N}$ is the data size; $n_{\mu}$ is the number of $[(\mu|\mu)_1,(\mu|\mu)_2,\dots,(\mu|\mu)_{N-1}]$ coincidence; $\phi_\mu^z$ is the phase error rate; $H_2(x)=-x\log_2x-(1-x)\log_2(1-x)$ is the binary Shannon entropy function;  $f$  is the error correction efficiency; $E_{\mu}^{1,j},j=2,3,\dots,N$, are marginal error rates, which describe the bit-flip error rates between users $U_1$ and $U_j$.  $\varepsilon _{\rm EC}$ and $\varepsilon _{\rm PA}$ are security coefficients regarding the correctness and secrecy.

In the asymptotic limit, the above key rate can be rewritten as  
\begin{equation}\label{QCKAlength}
 \begin{aligned}
		R=& Q_{\mu}\left[1-H_2(\overline{\phi}_{\mu}^z)-f\times\max\limits_{j\geq 2}  H_2(E^{1,j}_{\mu})\right],  
  \end{aligned}
\end{equation}
where $Q_{\mu}$ is the efficiency of producing  coincidences  $[(\mu|\mu)_1,(\mu|\mu)_2,\dots,(\mu|\mu)_{N-1}]$ in each time bin. We note that, since post-measurement coincidence matching is employed and there is no concept of a total sent match number, the terms ‘gain’ and ‘yield’ do not apply in our protocol. In the experiment, $s_\mu$ and $E_\mu^{1,j}$ can be measured directly, whereas the phase error rate $\phi_\mu^z$ is estimated by the multiparty decoy-state method.
 The phase error rate is determined by different photon-number components:
\begin{equation}
\begin{aligned}
\phi_{\mu}^z=\begin{cases}
\sum_{n=0}^\infty\frac{ s_{\mu}^{2n}}{s_{\mu}}=1-\sum_{n=0}^\infty \frac{s_{\mu}^{2n+1}}{s_{\mu}} \leq	1-\sum_{n=0}^{(N-2)/2} \frac{\underline{s}_{\mu}^{2n+1}}{s_{\mu}} &\mbox{if } N\in \rm even,\\
\sum_{n=0}^\infty\frac{ s_{\mu}^{2n+1}}{s_{\mu}}=1-\sum_{n=0}^\infty \frac{s_{\mu}^{2n}}{s_{\mu}} \leq	1-\sum_{n=0}^{(N-1)/2} \frac{\underline{s}_{\mu}^{2n}}{s_{\mu}} &\mbox{if } N\in \rm odd,
\end{cases}
\end{aligned} \label{phase-error}
\end{equation}
where $s_{\mu}^n$ is the number of contributions to the coincidences $[(\mu |\mu )_1,(\mu |\mu )_2,\dots,(\mu |\mu )_{N-1}]$ when the total number of photons emitted is $n$. 

\subsection{Simulation formulas}
We use the notation $(k_j|k_{j+1})_j$ to represent a successful click at the $j$-th measuring port, where users $U_j$ and $U_{j+1}$ respectively send states with intensities $k_j$ and $k_{j+1}$. The random phase sent by user $U_j$ is denoted as $\theta_j=2\pi\mathcal{M}_j/\mathcal{M}$ with  $\mathcal{M}_j\in\{0,1,\dots,\mathcal{M}-1\}$.
When users $U_j$ and $U_{j+1}$ send intensities $k_j$ and $k_{j+1}$ with a phase difference of $\Delta\theta$, the gain resulting from only detector $D_j^L$ or $D_j^R$ clicking is
\begin{equation}
\begin{aligned}
q_{(k_{j}|k_{j+1})}^{\Delta\theta}=  y_{(k_{j}|k_{j+1})}\left[e^{\eta_t
\sqrt{ k_{j}k_{j+1}}\cos{(\Delta\theta)}} +e^{-\eta_t\sqrt{ k_{j}k_{j+1}}\cos{(\Delta\theta)}}-2y_{(k_{j}|k_{j+1})}\right],
\end{aligned} 
\end{equation} 
 where $y_{(k_{j}|k_{j+1})}=\left(1-p_d \right)e^{-\frac{\eta_t \left(  k_{j}+ k_{j+1}\right)}{2}}$; $\eta_t=\frac{\eta_d}{2}10^{-\frac{\alpha L}{10}}$ is the total efficiency containing transmittance of channel, the transmittance of BS0 $(\frac{1}{2})$, and the efficiency of detector $\eta_d$; $p_d$ is the dark count rate of the detectors; $L$ is the distance between each user and Eve. By integrated the $\Delta\theta$ from 0 to $2\pi$, the overall gain  can be expressed as 
\begin{equation}
\begin{aligned}
q_{(k_{j}|k_{j+1})}&=\frac{1}{2\pi} \int_0^{2\pi}
q_{(k_{j}|k_{j+1})}^{\Delta\theta} d(\Delta\theta)  =2y_{(k_{j}|k_{j+1})}I_0\left(\eta_t\sqrt{ k_{j} k_{j+1}}\right) -2y_{(k_{j}|k_{j+1})}^2,
\end{aligned} 
\end{equation} 
where $I_0(x)$ refers to the zero-order modified Bessel function of the first kind. 

The number of retained successful time bins corresponding to click $(k|k)_j$, which satisfies $m=\mathcal{M}_j \equiv  
\mathcal{M}_{j+1}\pmod  {\frac{\mathcal{M}}{2}}$, is
\begin{equation}
\begin{aligned}
n_{(k |k )_j}^{m} =&\frac{4\mathcal{N}}{\mathcal{M}^2}
\sum\limits_{k_{w_1}}\sum\limits_{k_{w_2}}\dots\sum\limits_{k_{w_{N-2}}} \left\{\sum\limits_{ \tilde{v}= \mathcal{V}_j }
 p_{k_{w_1}}p_{k_{w_2}} \dots p_{k_{w_{N-2}}}(p_{k})^2q_{(k|k)}^{0} \prod_{i^\prime=1 }^{N-2}\left[1-q_{(k_{\tilde{v}_{i^\prime}}|k_{\tilde{v}_{i^\prime}+1})}\right]\right.\\
&+\left.\sum\limits_{l=1}^{N-2}\sum\limits_{\substack{v\in \mathcal{V}_{j,l}\\\tilde{v}= \mathcal{V}_j \setminus v }}
\frac{p_{k_{w_1}}p_{k_{w_2}} \dots p_{k_{w_{N-2}}}(p_{k})^2q_{(k|k)}^{0} }{l+1}\prod_{i=1 }^{l}q_{(k_{v_i}|k_{v_i+1})}\prod_{i^\prime=1 }^{N-l-2}\left[1-q_{(k_{\tilde{v}_{i^\prime}}|k_{\tilde{v}_{i^\prime}+1})}\right]\right\}\\
 =&\frac{4\mathcal{N}(p_{k})^2q_{(k|k)}^{0} }{\mathcal{M}^2} 
\sum\limits_{k_{w_1}}\sum\limits_{k_{w_2}}\dots\sum\limits_{k_{w_{N-2}}}  p_{k_{w_1}}p_{k_{w_2}} \dots p_{k_{w_{N-2}}}\left(1+\sum\limits_{l=1}^{N-2}\sum\limits_{v\in \mathcal{V}_{j,l}}\frac{(-1)^{l} }{l+1}\prod_{i=1 }^lq_{(k_{v_i}|k_{v_i+1})}\right),
\end{aligned} \label{number_click}
\end{equation}
where $w_i$ denotes $i$-th element of the set $\{1,2\dots, N\}\setminus\{j,j+1\}$ for $i=1$ to $N-2$, $\mathcal{V}_{j}=\{1,2\dots, N-1\}\setminus\{j\}$, and $\mathcal{V}_{j,l}=\{A|A\subseteq \mathcal{V}^j, |A|=l\}$. In the first equality, $l$ represents the number of additional measuring nodes that achieve successful clicks, with the probability that the $j$-th node is selected and retained being $\frac{1}{l+1}$. We observe that, as expected, the value of $n_{(k |k )_j}^{m}$ of do not depend on $m$.
The total number of set $\mathcal{T}_j^m$ is
 \begin{equation}
\begin{aligned}
 n^{m}_{j}= \sum\limits_{k\in \mathcal{K} } n_{(k |k )_j}^{m}.
\end{aligned} 
\end{equation}
where  $\mathcal{K}=\{\mu,\mu_1,\mu_2,\cdots,\mu_{N-1},0\}$. After post-matching, the total number of set $\mathcal{S}_{k}$ is
 \begin{equation}
\begin{aligned}
s_{ k}=& \sum_{m=0}^{\mathcal{M}/2-1} n_{\rm min}^{m} \prod_{j=1}^{N-1} \frac{n_{(k|k)_j}^{m}}{n_{j}^{m}}= \frac{\mathcal{M}}{2} n_{\rm min}^{0} \prod_{j=1}^{N-1} \frac{n_{(k|k)_j}^{0}}{n_{j}^{0}}, 
\end{aligned} 
\end{equation}
where $n_{\rm min}^{m}= \min\{n^{m}_{1},n^{m}_{2},\cdots,n^{m}_{ N-1 }\}$, and we use the fact that $n_{(k |k )_j}^{0}= n_{(k |k )_j}^{m}$ and $n_j^0=n_j^m$ for  all $m$ in the second equality. The bit error rate between users $U_j$ and $U_{j+1}$ can be written as 
 \begin{equation}
\begin{aligned}
E_{\mu}^{j,j+1} = \frac{q_{(\mu|\mu)}^{0, R}}{q_{(\mu|\mu)}^{0}}=\frac{e^{-\eta_t\sqrt{ k_{j}k_{j+1}}\cos{\theta}}-y_{(k_{j}|k_{j+1})}}{e^{\eta_t
\sqrt{ k_{j}k_{j+1}}\cos{\theta}} +e^{-\eta_t\sqrt{ k_{j}k_{j+1}}\cos{\theta}}-2y_{(k_{j}|k_{j+1})}}, 
\end{aligned} 
\end{equation}
where $q_{(\mu|\mu)}^{0, R}$ is the probability that only detector $D_j^R$ clicks when users $U_j$ and $U_{j+1}$ both send intensities $\mu$ with the same phase. Due to the symmetry of the channel mode, $E_{\mu}^{j,j+1}$ is independent of $j$ and can be labeled as $E_\mu$ for simplicity. The marginal error rate between users $U_1$ and $U_j$ can be estimated as
\begin{equation}
\begin{aligned}
E^{1,j}_{\mu}=\sum_{i=0}^{\lfloor(j-2)/2\rfloor}\binom{j-1}{2i+1} (E_{\mu})^{2i+1} (1-E_{\mu})^{j-2i-2}, 
\end{aligned} 
\end{equation}
where $\tbinom{j-1}{2i+1}=\frac{(j-1)!}{(2i+1)!(j-2i-2)!}$ is the binomial coefficient.

\section{Decoy-state estimation}
In the MF-QCKA protocol, the total intensity information for each coincidence match is random and independent of Eve's operation,  allowing decoy states to estimate the phase error rate $\phi_\mu^z$. For coincidences $[(k|k)_1,(k|k)_2,\dots,(k|k)_{N-1}]$, with $k\in\{\mu,\mu_1,\mu_2,\cdots,\mu_{N-1},0\}$, the phases sent at corresponding time bins are either identical or differ by $\pi$.  For simplicity, we assume that at these time bins, all users send the same phase, $\theta$. The phase $\theta$ is uniformly random and density matrix of the joint state can be written as
\begin{equation}
\begin{aligned}
\rho=\frac{1}{2\pi}\int_{0}^{2\pi}\bigotimes_{j=1}^N\ket{e^{i\theta}\sqrt{k}}_{u_j}\bra{e^{i\theta}\sqrt{k}}\bigotimes_{j=2}^{N-1}\ket{e^{i\theta}\sqrt{k}}_{\tilde{u}_j}\bra{e^{i\theta}\sqrt{k}}d\theta= e^{-2(N-1)k}\sum_n^\infty \frac{[2(N-1)k]^n}{n!}\ket{n}\bra{n},
\end{aligned} 
\end{equation}
where $\ket{n}$ denotes a Fock state with $n$ photons.  It is equivalent to preparing a virtual source $\ket{e^{i\theta} \sqrt{2(N-1)k}}$ with a total intensity of $2(N-1)k$ and a randomized phase $\theta$. We define $x^*$ as the expected value of $x$. When this virtual source emits exactly $n$ photons, the expected ratio of different coincidences should be the same as the ratio of the emitted states:
\begin{equation}
\begin{aligned}
\frac{s_k^{n*}}{s_{k'}^{n*}}=\dfrac{ (p_{k})^{2(N-1)}e^{-2(N-1)k} \frac{[2(N-1)k]^n}{n!}}{ (p_{{k'}})^{2(N-1)} e^{-2(N-1){k'}} \frac{[2(N-1){k'}]^n}{n!}},
\end{aligned} \label{number_relation}
\end{equation}
where $p_{k}$ is the probability of sending a state with intensity $k$ in each time bin.

\subsection{Decoy-state estimation of 3-user MF-QCKA}
For 3-user MF-QCKA,  the upper bound of phase error rate is
\begin{equation}
\begin{aligned}
\overline{\phi}_{\mu}^z
=1-\frac{\underline{s}_\mu^0+\underline{s}_\mu^2}{s_{\mu}}. 
\end{aligned} 
\end{equation}
In the case of 3-user MF-QCKA with three decoy states, $\nu$, $\omega$, and $o$, where $\mu>\nu>\omega>o=0$, we can extract lower bounds for the expected values of $s_\mu^0$ and $s_\mu^2$ using the decoy-state method~\cite{hwang2003decoy,lo2005decoy,wang2005decoy}. We combine the result from Eq.~\eqref{number_relation} with the formula $s_{k}^*=\sum_{n=0}^\infty s_{k}^{n*}$ to derive
\begin{equation}
\begin{aligned}
\frac{(p_\mu)^4s^*_o}{e^{4\mu}(p_o)^4}&=s_\mu^{0*}\\
s^*_{\mu}&=\sum_{n=0}^\infty s_\mu^{n*},\\
\frac{e^{4\nu}(p_\mu)^4s^*_{\nu}}{e^{4\mu}(p_\nu)^4}&=\sum_{n=0}^\infty\frac{\nu^n}{\mu^n}s^{n*}_\mu,\\
\frac{e^{4\omega}(p_\mu)^4s^*_{\omega}}{e^{4\mu}(p_\omega)^4}&=\sum_{n=0}^\infty\frac{\omega^n}{\mu^n}s^{n*}_\mu.
\end{aligned}
\end{equation}
We can cancel out the terms $s_\mu^{0*}$ and $s_\mu^{1*}$  using the Gaussian elimination method~\cite{grasselli2019practical} and generate equations given by 
\begin{equation}
\begin{aligned}
\nu\frac{(p_{\mu})^4}{e^{4\mu}}\left[e^{4\mu}\frac{s^*_{\mu}}{(p_\mu)^4}-\frac{s^*_o}{(p_o)^4}\right]-\mu\frac{(p_{\mu})^4}{e^{4\mu}} \left[e^{4\nu}\frac{s^*_{\nu}}{(p_\nu)^4}-\frac{s^*_o}{(p_o)^4}\right]&=\nu\mu\sum_{n=2}^\infty \frac{(\mu^{n-1}-\nu^{n-1})}{\mu^n}s_\mu^{n*},\\
\omega \frac{(p_{\mu})^4}{e^{4\mu}}\left[e^{4\nu}\frac{s^*_{\nu}}{(p_\nu)^4}-\frac{s^*_o}{(p_o)^4}\right]-\nu \frac{(p_{\mu})^4}{e^{4\mu}}\left[e^{4\omega}\frac{s^*_{\omega}}{(p_\omega)^4}-\frac{s^*_o}{(p_o)^4}\right]&=\omega\nu \sum_{n=2}^\infty \frac{(\nu^{n-1}-\omega^{n-1})}{\mu^n}s_\mu^{n*}.\\
\end{aligned}
\end{equation}
Then, we can further cancel out the term $s_\mu^{3*}$ as follows:
\begin{equation}
\begin{aligned}
\mu(\mu^2-\nu^2)a_2-\omega(\nu^2-\omega^2)a_1=&\mu\nu\omega\left[\frac{(\mu-\nu)(\nu-\omega)(\mu-\omega)}{\mu^2}\right]s_\mu^2\\
-&\sum_{n=4}^\infty \mu\nu\omega\left[\frac{(\mu^{n-1}-\nu^{n-1})(\nu^2-\omega^2)-(\nu^{n-1}-\omega^{n-1})(\mu^2-\nu^2)}{\mu^n}\right]s_\mu^{n*},
\end{aligned}\label{3user_eq}
\end{equation}
where $a_1^*=\nu\frac{(p_{\mu})^4}{e^{4\mu}}\left[e^{4\mu}\frac{s^*_{\mu}}{(p_\mu)^4}-\frac{s^*_o}{(p_o)^4}\right]-\mu\frac{(p_{\mu})^4}{e^{4\mu}} \left[e^{4\nu}\frac{s^*_{\nu}}{(p_\nu)^4}-\frac{s^*_o}{(p_o)^4}\right]$, $a_2^*=\omega \frac{(p_{\mu})^4}{e^{4\mu}}\left[e^{4\nu}\frac{s^*_{\nu}}{(p_\nu)^4}-\frac{s^*_o}{(p_o)^4}\right]-\nu \frac{(p_{\mu})^4}{e^{4\mu}}\left[e^{4\omega}\frac{s^*_{\omega}}{(p_\omega)^4}-\frac{s^*_o}{(p_o)^4}\right]$.
For  $n \ge 4$,
\begin{equation}
\begin{aligned}
&(\mu^{n-1}-\nu^{n-1})(\nu^2-\omega^2)-(\nu^{n-1}-\omega^{n-1})(\mu^2-\nu^2)=
(\mu- \nu) (\nu - \omega)\left[\sum_{i=0}^{n-2}\mu^{n-2-i}\nu^{i}(\nu + \omega)-\sum_{i=0}^{n-2}\omega^{n-2-i}\nu^{i}(\mu + \nu)\right]\\
&=(\mu- \nu) (\nu - \omega) (\mu- \omega)
\left[\sum_{j=0}^{n-3-i}\sum_{i=0}^{n-3}\mu^{n-3-i-j}\omega^{j}\nu^{i+1} + (\mu+\omega)\sum_{j=0}^{n-3-i}\sum_{i=0}^{n-3}\mu^{n-3-i-j}\omega^j\nu^{i}-\sum_{j=0}^{n-2-i}\sum_{i=0}^{n-2}\mu^{n-2-i-j}\omega^j\nu^{i}\right]\\
&=(\mu- \nu) (\nu - \omega) (\mu- \omega)
\left[\sum_{j=0}^{n-3-i}\sum_{i=0}^{n-3}\mu^{n-3-i-j}\omega^{j}\nu^{i+1} + \sum_{j=0}^{n-3-i}\sum_{i=0}^{n-3}\mu^{n-3-i-j}\omega^{j+1}\nu^{i}-\sum_{i=0}^{n-3}\omega^{n-2-i}\nu^{i}-\nu^{n-2}\right]\\
&=(\mu- \nu) (\nu - \omega) (\mu- \omega)
\left[\sum_{j=0}^{n-3-i}\sum_{i=0}^{n-4}\mu^{n-3-i-j}\omega^{j}\nu^{i+1} + \sum_{j=0}^{n-4-i}\sum_{i=0}^{n-4}\mu^{n-3-i-j}\omega^{j+1}\nu^{i}\right] > 0.
\end{aligned} \label{3user_ueq}
\end{equation}
Thus, by using Eq.~\eqref{3user_ueq} in Eq.~\eqref{3user_eq}, we can obtain the lower bound of $s_\mu^{2*}$:
\begin{equation}
\begin{aligned}
\underline{s}_\mu^{2*}=&\frac{\mu[\mu(\mu^2-\nu^2)a_2^*-\omega(\nu^2-\omega^2)a_1^*]}{\nu\omega(\mu-\nu)(\nu-\omega)(\mu-\omega)}.
\end{aligned} 
\end{equation}
In the limit of infinite data size, we have that $s_k=s_k^{*}$ and $s_k^n=s_k^{n*}$, and hence  
\begin{equation}
\begin{aligned}
\underline{s}_\mu^0=&s_\mu^0=\frac{(p_\mu)^4s_o}{e^{4\mu}(p_o)^4}\\
\underline{s}_\mu^2=&\frac{(p_\mu)^4\mu}{e^{4\mu}\nu\omega(\mu-\nu)(\nu-\omega)(\mu-\omega)}\times\left[ \frac{(\mu - \nu)(\nu -\omega)(\mu - \omega)  (\mu + \nu + \omega) s_o }{(p_o)^4}\right.\\
&+\left. \frac{
    e^{4 \nu} \mu\omega (\mu^2- \omega^2) s_\nu }{
    (p_\nu)^4} -\frac{e^{4 \omega } \mu \nu (\mu^2 - \nu^2) s_\omega}{(p_\omega)^4} -
 \frac{e^{4 \mu} \nu\omega  (\nu^2- \omega^2)s_\mu}{(p_\mu)^4} \right].
\end{aligned} 
\end{equation}

In the finite-size regime,
for three-party MF-QCKA,  we have
\begin{equation}
\begin{aligned}
\underline{s}_\mu^{0*}=&\frac{(p_\mu)^4\underline{s}^*_o}{e^{4\mu}(p_o)^4},\\
\underline{s}_\mu^{2*}=&
 \frac{(p_\mu)^4\mu}{e^{4\mu}\nu\omega(\mu-\nu)(\nu-\omega)(\mu-\omega)}\times\left[\frac{(\mu - \nu)(\nu -\omega)(\mu - \omega)  (\mu + \nu + \omega) \underline{s}^*_o }{(p_o)^4}\right.\\
&\left. +\frac{
    e^{4 \nu} \mu\omega (\mu^2- \omega^2) \underline{s}^*_\nu }{
    (p_\nu)^4} -\frac{e^{4 \omega } \mu \nu (\mu^2 - \nu^2) \overline{s}^*_\omega}{(p_\omega)^4} -
 \frac{e^{4 \mu} \nu\omega  (\nu^2- \omega^2)\overline{s}^*_\mu}{(p_\mu)^4} \right].
\end{aligned} \label{3user}
\end{equation}
 This equation is represented by expected values, but the values we get in  experiments are observed values. Using the variant of the Chernoff bound, we can obtain the upper and lower bounds of the expected value $s_k^*$  by~\cite{yin2020tight}:
\begin{equation}
\begin{aligned}\label{varchernoff1}	\overline{s}_k^{*}&=s_k+\beta+\sqrt{2\beta s_k+\beta^{2}}, \qquad 
\underline{s}_k^{*}=\max\left\{s_k-\frac{\beta}{2}-\sqrt{2\beta s_k+\frac{\beta^{2}}{4}},~0\right\},
\end{aligned}
\end{equation}
where $\beta=\ln{\epsilon^{-1}}$ and $\epsilon$ is failure probability.  Additionally, the Chernoff bound can help us estimate the lower bounds of $s_\mu^0$ and $s_\mu^2$ from their expected values~\cite{yin2020tight}:
\begin{equation}
\begin{aligned} 
\underline{s}_\mu^0 =\max\left\{\underline{s}_\mu^{0*}-\sqrt{2\beta \underline{s}_\mu^{0*}},~0\right\},\qquad
\underline{s}_\mu^2 =\max\left\{\underline{s}_\mu^{2*}-\sqrt{2\beta\underline{s}_\mu^{2*}}, ~0\right\}. 
\end{aligned}
\end{equation}

\subsection{Decoy-state estimation of 4-user MF-QCKA}
For 4-user MF-QCKA with four decoy states, $\nu$, $\omega$, $\xi$, and $o$, where $\mu>\nu>\omega>\xi>o=0$, as defined in Eq.~\eqref{phase-error} and by applying the result in Eq.~\eqref{number_relation}, we have
\begin{equation}
\begin{aligned}
\overline{\phi}_{\mu}^z
=	1-\frac{\underline{s}_\mu^1+\underline{s}_\mu^3}{s_{\mu}},
\end{aligned} 
\end{equation}
and
\begin{equation}
\begin{aligned}
\frac{(p_\mu)^6s^*_o}{e^{6\mu}(p_o)^6}&=s_\mu^{0*},  \\
s^*_{\mu}&=\sum_{n=0}^\infty s^{n*}_\mu,\\
\frac{e^{6\nu}(p_\mu)^6s^*_{\nu}}{e^{6\mu}(p_\nu)^6} &=\sum_{n=0}^\infty\frac{\nu^n}{\mu^n}s^{n*}_\mu,\\
\frac{e^{6\omega}(p_\mu)^6s^*_{\omega}}{e^{6\mu}(p_\omega)^6}&=\sum_{n=0}^\infty\frac{\omega^n}{\mu^n}s^{n*}_\mu,\\
\frac{e^{6\xi}(p_\mu)^6s^*_{\xi}}{e^{6\mu}(p_\xi)^6}&=\sum_{n=0}^\infty\frac{\xi^n}{\mu^n}s^{n*}_\mu.
\end{aligned}
\end{equation}
Again, in the limit of infinite data size,  $s_k=s_k^*$ and $s_k^n=s_k^{n*}$.  This allows us to use the decoy-state method to calculate the lower bound of $s_\mu^1$, which is
\begin{equation}
\begin{aligned}
\underline{s}_\mu^1&=\frac{\mu\omega^2\frac{(p_{\mu})^6}{e^{6\mu}} \left[e^{6\xi}\frac{s_{\xi}}{(p_\xi)^6}-\frac{s_o}{(p_o)^6}\right]-\mu\xi^2 \frac{(p_{\mu})^6}{e^{6\mu}} \left[e^{6\omega}\frac{s_{\omega}}{(p_\omega)^6}-\frac{s_o}{(p_o)^6}\right]}{\xi\omega(\omega-\xi)}\\
&=\frac{\mu\omega (p_{\mu})^6e^{-6\mu}}{(\omega\xi-\xi^2) } \left[e^{6\xi}\frac{s_{\xi}}{(p_\xi)^6} - \frac{\xi^2}{\omega^2} e^{6\omega}\frac{s_{\omega}}{(p_\omega)^6}-\frac{\omega^2-\xi^2}{\omega^2} \frac{s_o}{(p_o)^6}\right].
\end{aligned} 
\end{equation}
To obtain a tight lower bound for $s_\mu^{3*}$, we need to cancel out the terms $s_\mu^{0*}$, $s_\mu^{1*}$ and $s_\mu^{2*}$  using the Gaussian elimination method, generating equations given by 
\begin{equation}
\begin{aligned}
\nu\frac{(p_{\mu})^6}{e^{6\mu}}  \left[e^{6\mu}\frac{s^*_{\mu}}{(p_\mu)^6}-\frac{s^*_o}{(p_o)^6}\right]-\mu \frac{(p_{\mu})^6}{e^{6\mu}} \left[e^{6\nu}\frac{s^*_{\nu}}{(p_\nu)^6}-\frac{s^*_o}{(p_o)^6}\right]&=\nu\mu\sum_{n=2}^\infty \frac{(\mu^{n-1}-\nu^{n-1})}{\mu^n}s_\mu^{n*},\\
\omega \frac{(p_{\mu})^6}{e^{6\mu}} \left[e^{6\nu}\frac{s^*_{\nu}}{(p_\nu)^6}-\frac{s^*_o}{(p_o)^6}\right]-\nu \frac{(p_{\mu})^6}{e^{6\mu}} \left[e^{6\omega}\frac{s^*_{\omega}}{(p_\omega)^6}-\frac{s^*_o}{(p_o)^6}\right]&=\omega\nu\sum_{n=2}^\infty \frac{(\nu^{n-1}-\omega^{n-1})}{\mu^n}s_\mu^{n*},\\
\xi \frac{(p_{\mu})^6}{e^{6\mu}} \left[e^{6\omega}\frac{s^*_{\omega}}{(p_\omega)^6}-\frac{s^*_o}{(p_o)^6}\right]-\omega\frac{(p_{\mu})^6}{e^{6\mu}} \left[e^{6\xi}\frac{s^*_{\xi}}{(p_\xi)^6}-\frac{s^*_o}{(p_o)^6}\right]&=\xi\omega\sum_{n=2}^\infty \frac{(\omega^{n-1}-\xi^{n-1})}{\mu^n}s_\mu^{n*}.\\
\end{aligned}
\end{equation}  
and 
\begin{equation}
\begin{aligned}
a^*_2\mu(\mu-\nu)-a^*_1\omega(\nu-\omega)&=\sum_{n=3}^\infty\frac{\nu\mu\omega[(\nu^{n-1}-\omega^{n-1})(\mu-\nu)-(\mu^{n-1}-\nu^{n-1})(\nu-\omega)]}{\mu^n}s_\mu^{n*},\\
a^*_3\nu(\nu-\omega)-a^*_2\xi(\omega-\xi)&=\sum_{n=3}^\infty\frac{\omega\nu\xi[(\omega^{n-1}-\xi^{n-1})(\nu-\omega)-(\nu^{n-1}-\omega^{n-1})(\omega-\xi)]}{\mu^n}s_\mu^{n*},\\
\end{aligned}
\end{equation}
where $a^*_1=\nu\frac{(p_{\mu})^6}{e^{6\mu}}  \left[e^{6\mu}\frac{s^*_{\mu}}{(p_\mu)^6}-\frac{s^*_o}{(p_o)^6}\right]-\mu \frac{(p_{\mu})^6}{e^{6\mu}} \left[e^{6\nu}\frac{s^*_{\nu}}{(p_\nu)^6}-\frac{s^*_o}{(p_o)^6}\right]$, $a^*_2=\omega \frac{(p_{\mu})^6}{e^{6\mu}} \left[e^{6\nu}\frac{s^*_{\nu}}{(p_\nu)^6}-\frac{s^*_o}{(p_o)^6}\right]-\nu \frac{(p_{\mu})^6}{e^{6\mu}} \left[e^{6\omega}\frac{s^*_{\omega}}{(p_\omega)^6}-\frac{s^*_o}{(p_o)^6}\right]$, $a^*_3=\xi \frac{(p_{\mu})^6}{e^{6\mu}} \left[e^{6\omega}\frac{s^*_{\omega}}{(p_\omega)^6}-\frac{s^*_o}{(p_o)^6}\right]-\omega\frac{(p_{\mu})^6}{e^{6\mu}} \left[e^{6\xi}\frac{s^*_{\xi}}{(p_\xi)^6}-\frac{s^*_o}{(p_o)^6}\right]$. Then, we can further cancel out the term $s_\mu^{4*}$ as follows:
\begin{equation}
\begin{aligned}
&b^*_1\xi (\omega-\xi)(\nu-\xi)(\xi+\omega+\nu)-b^*_2\mu(\mu - \nu) (\mu - \omega)   (\mu + \nu + \omega)\\
=&\frac{\nu\mu\omega\xi(\mu-\nu)(\mu-\xi)(\nu-\xi)(\nu - \omega) (\mu - \omega)   (\omega-\xi)}{\mu^3}s_\mu^{3*}\\
-&\sum_{n=5}^\infty\frac{\nu\mu\omega\xi(\mu-\nu)(\mu-\xi)(\nu-\xi)(\nu - \omega) (\mu - \omega)   (\omega-\xi)\varOmega_n}{\mu^n}s_\mu^{n*},\label{eq_use4}
\end{aligned}
\end{equation}
where $b^*_1=a^*_2\mu(\mu-\nu)-a^*_1\omega(\nu-\omega)$, $b^*_2=a^*_3\nu(\nu-\omega)-a^*_2\xi(\omega-\xi)$, and 
\begin{equation}
\begin{aligned}
\varOmega_n= &[(\mu^{n-1}-\nu^{n-1})(\nu-\omega)-(\nu^{n-1}-\omega^{n-1})(\mu-\nu)]  (\omega-\xi)(\nu-\xi)(\xi+\omega+\nu) \\
&- [(\nu^{n-1}-\omega^{n-1})(\omega-\xi)-(\omega^{n-1}-\xi^{n-1})(\nu-\omega)](\mu - \nu) (\mu - \omega)   (\mu + \nu + \omega)\\
=&(\mu-\nu)(\nu-\omega)(\mu-\omega)(\omega-\xi)(\nu-\xi)\sum_{j=0}^{n-3-i}\sum_{i=0}^{n-3}\mu^{n-3-i-j}\omega^{j}\nu^{i}(\nu+\omega+\xi)\\
&-(\nu-\omega)(\omega-\xi)(\nu-\xi)(\mu-\nu)(\mu-\omega) \sum_{j=0}^{n-3-i}\sum_{i=0}^{n-3}\nu^{n-3-i-j}\xi^{j}\omega^{i}(\mu+\nu+\omega)\\
=&(\mu-\nu)(\nu-\omega)(\mu-\omega)(\omega-\xi)(\nu-\xi)\\
&\times\left[(\nu+\omega+\xi)\sum_{j=0}^{n-3-i}\sum_{i=0}^{n-3}\mu^{n-3-i-j}\omega^{j}\nu^{i}- (\mu+\nu+\omega)\sum_{j=0}^{n-3-i}\sum_{i=0}^{n-3}\xi^{n-3-i-j}\omega^{j}\nu^{i}\right]\\
=&(\mu-\nu)(\nu-\omega)(\mu-\omega)(\omega-\xi)(\nu-\xi)\\
&\times\left[(\nu+\omega)\sum_{j=0}^{n-3-i}\sum_{i=0}^{n-3}\left(\mu^{n-3-i-j}-\xi^{n-3-i-j}\right)\omega^{j}\nu^{i}+ (\mu+\xi)\sum_{j=0}^{n-3-i}\sum_{i=0}^{n-3}\left(\mu^{n-3-i-j}-\xi^{n-3-i-j}\right)\omega^{j}\nu^{i}\right.\\
&-\left.\sum_{j=0}^{n-3-i}\sum_{i=0}^{n-3}\left(\mu^{n-2-i-j}-\xi^{n-2-i-j}\right)\omega^{j}\nu^{i}\right]\\
=&(\mu-\nu)(\nu-\omega)(\mu-\omega)(\omega-\xi)(\nu-\xi)(\mu-\xi)\\
&\times\left[(\nu+\omega)\sum_{k=0}^{n-4-i-j}\sum_{j=0}^{n-4-i}\sum_{i=0}^{n-4}\mu^{n-4-i-j-k}\xi^{k}\omega^{j}\nu^{i}+ (\mu+\xi)\sum_{k=0}^{n-4-i-j}\sum_{j=0}^{n-4-i}\sum_{i=0}^{n-4}\mu^{n-4-i-j-k}\xi^{k}\omega^{j}\nu^{i}\right.\\
&-\left.\sum_{k=0}^{n-3-i-j}\sum_{j=0}^{n-3-i}\sum_{i=0}^{n-3}\mu^{n-3-i-j-k}\xi^{k}\omega^{j}\nu^{i}\right]\\
=&(\mu-\nu)(\nu-\omega)(\mu-\omega)(\omega-\xi)(\nu-\xi)(\mu-\xi)\\
&\times\left[(\nu+\omega)\sum_{k=0}^{n-4-i-j}\sum_{j=0}^{n-4-i}\sum_{i=0}^{n-4}\mu^{n-4-i-j-k}\xi^{k}\omega^{j}\nu^{i}+ \sum_{k=0}^{n-4-i-j}\sum_{j=0}^{n-4-i}\sum_{i=0}^{n-4}\mu^{n-4-i-j-k}\xi^{k+1}\omega^{j}\nu^{i}\right.\\
&-\left.\sum_{j=0}^{n-3-i}\sum_{i=0}^{n-4}\xi^{n-3-i-j}\omega^{j}\nu^{i}-\nu^{n-3}\right]\\
=&(\mu-\nu)(\nu-\omega)(\mu-\omega)(\omega-\xi)(\nu-\xi)(\mu-\xi)\\
&\times\left[\sum_{k=0}^{n-4-i-j}\sum_{j=0}^{n-4-i}\sum_{i=0}^{n-5}\mu^{n-4-i-j-k}\xi^{k}\omega^{j}\nu^{i+1}+ \sum_{k=0}^{n-4-i-j}\sum_{j=0}^{n-4-i}\sum_{i=0}^{n-4}\omega^{n-3-i-j-k}\xi^{k}\mu^{j}\nu^{i}- \sum_{i=0}^{n-4}\omega^{n-3-i}\nu^{i}\right.\\
&+\left. \sum_{k=0}^{n-5-i-j}\sum_{j=0}^{n-5-i}\sum_{i=0}^{n-5}\mu^{n-4-i-j-k}\xi^{k+1}\omega^{j}\nu^{i}\right] >0.
\end{aligned}\label{proof_use4}
\end{equation}
The last inequality  is derived by  using the fact that $\mu>\nu>\omega>\xi>0$. Finally, by employing Eq.~\eqref{proof_use4} in Eq.~\eqref{eq_use4}, we obtain 
\begin{equation}
\begin{aligned}
\underline{s}_\mu^{3*}=\frac{\mu^2[b^*_1\xi (\omega-\xi)(\nu-\xi)(\xi+\omega+\nu)-b^*_2\mu(\mu - \nu) (\mu - \omega)   (\mu + \nu + \omega)]}{ \nu \omega\xi(\mu-\nu)(\mu-\xi)(\nu-\xi)(\nu - \omega) (\mu - \omega)   (\omega-\xi)}.
\end{aligned} 
\end{equation}
In the case of infinite size, we have
\begin{equation}
\begin{aligned}
\underline{s}_\mu^3&=\frac{\mu^2[b_1\xi (\omega-\xi)(\nu-\xi)(\xi+\omega+\nu)-b_2\mu(\mu - \nu) (\mu - \omega)   (\mu + \nu + \omega)]}{ \nu \omega\xi(\mu-\nu)(\mu-\xi)(\nu-\xi)(\nu - \omega) (\mu - \omega)   (\omega-\xi)},
\end{aligned} 
\end{equation}
where $b_1=a_2\mu(\mu-\nu)-a_1\omega(\nu-\omega)$, $b_2=a_3\nu(\nu-\omega)-a_2\xi(\omega-\xi)$, $a_1=\nu\frac{(p_{\mu})^6}{e^{6\mu}}  \left[e^{6\mu}\frac{s_{\mu}}{(p_\mu)^6}-\frac{s_o}{(p_o)^6}\right]-\mu \frac{(p_{\mu})^6}{e^{6\mu}} \left[e^{6\nu}\frac{s_{\nu}}{(p_\nu)^6}-\frac{s_o}{(p_o)^6}\right]$, $a_2=\omega \frac{(p_{\mu})^6}{e^{6\mu}} \left[e^{6\nu}\frac{s_{\nu}}{(p_\nu)^6}-\frac{s_o}{(p_o)^6}\right]-\nu \frac{(p_{\mu})^6}{e^{6\mu}} \left[e^{6\omega}\frac{s_{\omega}}{(p_\omega)^6}-\frac{s_o}{(p_o)^6}\right]$, and $a_3=\xi \frac{(p_{\mu})^6}{e^{6\mu}} \left[e^{6\omega}\frac{s_{\omega}}{(p_\omega)^6}-\frac{s_o}{(p_o)^6}\right]-\omega\frac{(p_{\mu})^6}{e^{6\mu}} \left[e^{6\xi}\frac{s_{\xi}}{(p_\xi)^6}-\frac{s_o}{(p_o)^6}\right]$. 

\subsection{Decoy-state estimation of 5-user MF-QCKA}
In a 5-user MF-QCKA setup, we use five decoy states: $\nu$, $\omega$, $\xi$, $\tau$, and $o$, with $\mu>\nu>\omega>\xi>\tau>o=0$. As defined in Eq.~\eqref{phase-error} and by applying the result from Eq.~\eqref{number_relation}, we have
\begin{equation}
\begin{aligned}
\overline{\phi}_{\mu}^z
=	1-\frac{\underline{s}_\mu^0+\underline{s}_\mu^2+\underline{s}_\mu^4}{s_{\mu}},
\end{aligned} 
\end{equation}
and
\begin{equation}
\begin{aligned}
\frac{(p_\mu)^8s^*_o}{e^{8\mu}(p_o)^8}&=  s^{0*}_\mu,\\
s^*_{\mu}&=\sum_{n=0}^\infty s^{n*}_\mu,\\
\frac{e^{8\nu}(p_\mu)^8s^*_{\nu}}{e^{8\mu}(p_\nu)^8}&=\sum_{n=0}^\infty\frac{\nu^n}{\mu^n}s^{n*}_\mu,\\
\frac{e^{8\omega}(p_\mu)^8s^*_{\omega}}{e^{8\mu}(p_\omega)^8}&=\sum_{n=0}^\infty\frac{\omega^n}{\mu^n}s^{n*}_\mu,\\
\frac{e^{8\xi}(p_\mu)^8s^*_{\xi}}{e^{8\mu}(p_\xi)^8}&=\sum_{n=0}^\infty\frac{\xi^n}{\mu^n}s^{n*}_\mu,\\
\frac{e^{8\tau}(p_\mu)^8s^*_{\tau}}{e^{8\mu}(p_\tau)^8}&= \sum_{n=0}^\infty\frac{\tau^n}{\mu^n}s^{n*}_\mu.
\end{aligned}
\end{equation}
Similarly, in the case of infinite size, and by using the Gaussian-elimination method, we obtain:
\begin{equation}
\begin{aligned}
\underline{s}_\mu^0&=s_\mu^0=\frac{(p_\mu)^8s_o}{e^{8\mu}(p_o)^8},\\
\underline{s}_\mu^2&=\frac{\mu^2[\tau(\xi^2-\tau^2)a_3-\omega(\omega^2-\xi^2)a_4]}{\omega\xi\tau(\omega-\xi)(\xi-\tau)(\omega-\tau)},\\
\underline{s}_\mu^4&=\frac{\mu^3[c_1\tau(\omega-\tau)(\xi-\tau)(\nu-\tau)(\nu+\omega+\xi+\tau)-c_2\mu(\mu-\nu)(\mu-\omega)(\mu-\xi)(\mu+\nu+\omega+\xi)]}{\omega\nu\xi\tau(\mu-\nu)(\mu-\omega)(\mu-\xi)(\nu-\omega)(\nu-\xi)(\omega-\xi)(\omega-\tau)(\xi-\tau)(\nu-\tau)(\mu-\tau)}, 
 \end{aligned}
\end{equation}
where   $a_1=\nu\frac{(p_{\mu})^8}{e^{8\mu}}  \left[e^{8\mu}\frac{s_{\mu}}{(p_\mu)^8}-\frac{s_o}{(p_o)^8}\right]-\mu \frac{(p_{\mu})^8}{e^{8\mu}} \left[e^{8\nu}\frac{s_{\nu}}{(p_\nu)^8}-\frac{s_o}{(p_o)^8}\right]$, $a_2=\omega \frac{(p_{\mu})^8}{e^{8\mu}} \left[e^{8\nu}\frac{s_{\nu}}{(p_\nu)^8}-\frac{s_o}{(p_o)^8}\right]-\nu \frac{(p_{\mu})^8}{e^{8\mu}} \left[e^{8\omega}\frac{s_{\omega}}{(p_\omega)^8}-\frac{s_o}{(p_o)^8}\right]$, $a_3=\xi \frac{(p_{\mu})^8}{e^{8\mu}} \left[e^{8\omega}\frac{s_{\omega}}{(p_\omega)^8}-\frac{s_o}{(p_o)^8}\right]-\omega\frac{(p_{\mu})^8}{e^{8\mu}} \left[e^{8\xi}\frac{s_{\xi}}{(p_\xi)^8}-\frac{s_o}{(p_o)^8}\right]$,
$a_4=\tau\frac{(p_\mu)^8}{e^{8\mu}}  \left[\frac{e^{8\xi}( s_{\xi}}{ (p_\xi)^8}-\frac{ s_o}{(p_o)^8}\right]-\xi\frac{(p_\mu)^8}{e^{8\mu}} \left[\frac{e^{8\tau} s_{\tau}}{ (p_\tau)^8}-\frac{ s_o}{ (p_o)^8}\right]$,
$b_1=a_2\mu(\mu-\nu)-a_1\omega(\nu-\omega)$, $b_2=a_3\nu(\nu-\omega)-a_2\xi(\omega-\xi)$, $b_3=a_4\omega(\omega-\xi)-a_3\tau(\xi-\tau)$, $c_1=b_1\xi (\omega-\xi)(\nu-\xi)-b_2\mu(\mu - \nu) (\mu - \omega) $, and $c_2=b_2\tau(\omega-\tau)(\xi-\tau)-b_3\nu(\nu-\omega)(\nu-\xi)$.

\section{Analytical expression for $s_{\mu}^n$}
Here we calculate the analytical expression for $s_{\mu}^n$ in Eq.~\eqref{phase-error}. From the definite of $s_{\mu}^n$,  we can write it as follows:
\begin{equation}
\begin{aligned}
s_{\mu}^n=&  \sum_{m=0}^{\mathcal{M}/2-1}\sum_{\mathcal{L}(n_1,\dots,n_{N-1})} n_{\rm min}^{m} \prod_{j=1}^{N-1} \frac{f_{n_j}^{m}}{n_{j}^{m}}, 
\end{aligned} \label{eq_number_conri}
\end{equation}
where  $\mathcal{L}(n_1,\dots,n_{N-1}):=\{ (n_1,\dots,n_{N-1}): 0\leq n_i \leq n,\sum_{i=1}^{N-1}n_i=n\}$, and $f_{n_j}^{m}$ denotes the number of retained click  $(\mu|\mu)_j$ that satisfies $m=\mathcal{M}_j \equiv  
\mathcal{M}_{j+1}\pmod  {\frac{\mathcal{M}}{2}}$ when users $U_j$ and $U_{j+1}$ transmit total photon numbers $n_j$.  With analogous calculations to those leading to Eq.~\eqref{number_click}, one can simplify $f_{n_j}^{m}$ as follows:
\begin{equation}
\begin{aligned}
f_{n_j}^{m}=&\frac{4\mathcal{N}(p_{\mu})^2}{\mathcal{M}^2} \sum_{l_j=0}^{n_j}\frac{ e^{-2\mu}(\mu)^{n_j}}{l_j!(n_j-l_j)!} Y_{(l_j,n_j-l_j)}\times\\
&\left[\sum\limits_{k_{w_1}}\sum\limits_{k_{w_2}}\dots\sum\limits_{k_{w_{N-2}}}  p_{k_{w_1}}p_{k_{w_2}} \dots p_{k_{w_{N-2}}}\left(1+\sum\limits_{l=1}^{N-2}\sum\limits_{v\in \mathcal{V}_{j,l}}\frac{(-1)^{l} }{l+1}\prod_{i=1 }^lq'_{(k_{v_i}|k_{v_i+1})}\right)\right],
\end{aligned}  
\end{equation}
where
\begin{equation}
\begin{aligned}
q'_{(k_{v_i}|k_{v_i+1})}
:=&\begin{cases}
q_{(k_{j-1}|\mu,l_j)},& v_i=j-1,\\
q_{(\mu,n_j-l_j|k_{j+2})},& v_i=j+1, \\
q_{(k_{v_i}|k_{v_i+1})},&\mbox{others}.
\end{cases}
\end{aligned}  
\end{equation}
where $Y_{(l_j,n_j-l_j)}$ denotes the probability of  click $(\mu|\mu)_{j}$ when users $U_j$ and $U_{j+1}$ respectively send $l_j$ and $n_j-l_j$ photons; $q_{(k_{j-1}|\mu,l_j)}$ is the probability of click $(k_{j-1}|\mu)_{j-1}$ given that user $U_j$ sent  $l_j$ photons, and $q_{(\mu, n_j-l_j|k_{j+1})}$ is the probability of click $(\mu|k_{j+1})_{j+1}$ given that user $U_{j+1}$ sent $n_j-l_j$ photons. To simplify the calculation, we take the first-order approximation to $f_{n_j}^{m}$ and obtain
\begin{equation}
\begin{aligned}
f_{n_j}^{m}\approx&\frac{4\mathcal{N}(p_{\mu})^2}{\mathcal{M}^2} \sum_{l_j=0}^{n_j}\frac{ e^{-2\mu}(\mu)^{n_j}}{l_j!(n_j-l_j)!} Y_{(l_j,n_j-l_j)}.
\end{aligned} 
\end{equation}
We remark that this approximation is a highly tight bound in high-loss regimes due to the low probability of additional measuring nodes achieving successful clicks ($l \ge 1$).

We now focus on the computation of $Y_{(l_j,n_j-l_j)}$. Let the compressed  notation $\ket{l_j,n_j-l_j}$  indicate $\ket{l_j}\ket{n_j-l_j}$. We combine the channel transmittance with the detector's efficiency, a widely applied approach in QKD simulations that does not affect the calculation. After going through the lossy channel and the beam splitter BS0, the state $\ket{l_j,n_j-l_j}\bra{l_j,n_j-l_j}$ evolves to
\begin{equation}
\begin{aligned}
\sum_{f_j=0}^{l_j}\sum_{g_j=0}^{n_j-l_j}\dbinom{l_j}{f_j}\dbinom{n_j-l_j}{g_j}  (\eta_t)^{f_j+g_j}(1-\eta_t)^{n_j-f_j-g_j} \ket{f_j,g_j}\bra{f_j,g_j},
\end{aligned} 
\end{equation}
where the loss mode is traced out since it couples to the environment. After the state $\ket{f_j,g_j}$ passes through BS, the output state reads
\begin{equation}
\begin{aligned}
\ket{\zeta(f_j,g_j)}&=\frac{ [(\hat{b}_{j}^\dagger+\hat{b}_{j,\bot}^\dagger)/\sqrt{2}]^{f_j}}{\sqrt{f_j!}}\frac{ [(\hat{b}_{j}^\dagger+\hat{b}_{j,\bot}^\dagger)/\sqrt{2}]^{g_j}}{\sqrt{g_j!}}\ket{0}\\
&= \sum_{f'_j=0}^{f_j}\sum_{g'_j=0}^{g_j}\dbinom{f_j}{f'_j}\dbinom{g_j}{g'_j}\frac{ (\hat{b}_{j}^\dagger)^{f'_j+g'_j}(\hat{b}_{j,\bot}^\dagger)^{f_j+g_j-f'_j-g'_j}}{\sqrt{2^{f_j+g_j}f_j!g_j!}} \ket{0}.\\
& = \sum_{f'_j=0}^{f_j}\sum_{g'_j=0}^{g_j}\frac{\sqrt{f_j!g_j! (f'_j+g'_j)!(f_j+g_j-f'_j-g'_j)!}}{\sqrt{2^{f_j+g_j}}f'_j!(f_j-f'_j)!g_j!(g_j-g'_j)!} \ket{f'_j+g'_j,f_j+g_j-f'_j-g'_j}\\
\end{aligned} 
\end{equation}
At this point, detectors $D_j^L$ and $D_j^R$  each perform a threshold measurement, returning a click in the corresponding detector if one or more photons are detected. We are interested in the probability that only detector $D_j^L$ or $D_j^R$ clicks. By including the effect of dark counts, we can express this probability as follows:
\begin{equation}
\begin{aligned}
P_{f_j,g_j}=&2p_d(1-p_d)\rm Tr\Big[\rho'\otimes\ket{0,0}\bra{0,0}\Big]+(1-p_d)Tr\left[\rho'\otimes\left(\sum_{n=1}^{\infty}\ket{n,0}\bra{n,0}+\sum_{n=1}^{\infty}\ket{0,n}\bra{0,n}\right)\right]\\
=&\begin{cases}
 2p_d(1-p_d),&\mbox{if } f_j+g_j=0,\\
\frac{ 2(1-p_d)(f_j+g_j)!} {2^{f_j+g_j}  f_j! g_j! },  &\mbox{if } f_j+g_j\neq 0.
\end{cases}
\end{aligned}  
\end{equation}
where $\rho'=\ket{\zeta(f_j,g_j)}\bra{\zeta(f_j,g_j)}$.
By  defining $\delta_0(x)$ such that $\delta_0(x):=1$ if $x=0$ and $\delta_0(x):=0$ if $x\neq 0$, we can abbreviate the above formula as $P_{f_j,g_j}=\frac{ 2(p_d)^{\delta_0(f_j+g_j)}(1-p_d)(f_j+g_j)!} {2^{f_j+g_j}  f_j! g_j! }$. Therefore, we have
\begin{equation}
\begin{aligned}
Y_{(l_j,n_j-l_j)}=  &\sum_{f_j=0}^{l_j}\sum_{g_j=0}^{n_j-l_j}\dbinom{l_j}{f_j}\dbinom{n_j-l_j}{g_j}  (\eta_t)^{f_j+g_j}(1-\eta_t)^{n_j-f_j-g_j} P_{f_j,g_j} \\
=&\sum_{f_j=0}^{l_j}\sum_{g_j=0}^{n_j-l_j}   \frac{2 (p_d)^{\delta_0(f_j+g_j)}(1-p_d)(\eta_t)^{f_j+g_j}(1-\eta_t)^{n_j-f_j-g_j} l_j!(n_j-l_j)!(f_j+g_j)!} {2^{f_j+g_j}(l_j-f_j)!(n_j-l_j-g_j)!  (f_j!)^2 (g_j!)^2}.
\end{aligned} 
\end{equation}
By using this expression in Eq.~\eqref{eq_number_conri}, we obtain
\begin{equation}
\begin{aligned}
s_{\mu}^n=&  \sum_{m=0}^{\mathcal{M}/2-1}\sum_{\mathcal{L}(n_1,\dots,n_{N-1})} n_{\rm min}^{m}  \prod_{j=1}^{N-1} \frac{1}{n_{j}^{m} }  \left[ \frac{4\mathcal{N}(p_{\mu})^2}{\mathcal{M}^2} \sum_{l_j=0}^{n_j}\frac{ e^{-2\mu}(\mu)^{n_j}}{l_j!(n_j-l_j)!} Y_{(l_j,n_j-l_j)}\right], \\
=&\frac{\mathcal{M}n_{\rm min}^{0}e^{-2(N-1)\mu}(\mu)^{n}(p_{\mu})^{2(N-1)}}{2\prod_{j=1}^{N-1} n_{j}^{0}} \sum_{\mathcal{L}(n_1,\dots,n_{N-1})}    \prod_{j=1}^{N-1}  \left[\sum_{l_j=0}^{n_j} \frac{4\mathcal{N}Y_{(l_j,n_j-l_j)}}{\mathcal{M}^2l_j!(n_j-l_j)!}    \right]\\
=&\frac{\mathcal{M}n_{\rm min}^{0}e^{-2(N-1)\mu}(\mu)^{n}(p_{\mu})^{2(N-1)}}{2\prod_{j=1}^{N-1} n_{j}^{0}}   \sum_{\mathcal{L}(n_1,\dots,n_{N-1})}\\
&\times \prod_{j=1}^{N-1} \left[\frac{4\mathcal{N}}{\mathcal{M}^2}\sum_{l_j=0}^{n_j}\sum_{f_j=0}^{l_j}\sum_{g_j=0}^{n_j-l_j}   \frac{2(p_d)^{\delta_0(f_j+g_j)}(1-p_d)(\eta_t)^{f_j+g_j}(1-\eta_t)^{n_j-f_j-g_j}   (f_j+g_j)!} {2^{f_j+g_j}(l_j-f_j)!(n_j-l_j-g_j)!  (f_j!)^2 (g_j!)^2}\right],
\end{aligned} \label{compu_number}
\end{equation}
where we use the fact that $n_{\rm min}^0=n_{\rm min}^m$ and $n_{j}^0=n_{j}^m$. By applying Eq.~\eqref{compu_number} to Eq.~\eqref{phase-error}, we derive a computable lower bound on the asymptotic conference key rate of our MF-QCKA protocol.

\newpage

%